\documentclass[%
 reprint,
 amsmath,amssymb,
 aps,
]{revtex4-2}

\usepackage{graphicx}
\usepackage{dcolumn}
\usepackage{bm}
\usepackage{hyperref}

\usepackage{enumitem}
\usepackage{braket}
\usepackage[caption=false]{subfig}

\begin{document}

\preprint{APS/123-QED}

\title{Scaling up prime factorization with self-organizing gates: A memcomputing approach}

\author{Tristan Sharp}
\author{Rishabh Khare}%
\author{Erick Pederson}%
\author{Fabio Lorenzo Traversa}
 \email{ftraversa@memcpu.com}
\affiliation{MemComputing, Inc. San Diego, California}%




\date{\today}

\begin{abstract}

We report preliminary results on using the MEMCPU\texttrademark{} Platform to compute the prime factorization of large biprimes. The first approach, the direct model, directly returns the factors of a given biprime. The second approach, the congruence model, returns smooth congruences to address the bottleneck of standard sieve methods. The models have size-dependent structure, and the MEMCPU Platform requires structure-dependent tuning for optimal performance. Therefore, for both models, we tuned the platform on sample problems up to a given size according to available resources. Then we generated RSA-like benchmark biprimes to perform rigorous scaling analysis. The MEMCPU timings over the tuned range followed low degree polynomials in the number of bits, markedly different than other tested methods including general number field sieve. MEMCPU's congruence model was the most promising, which was scaled up to 300-bit factorization problems while following a $2^{nd}$ degree polynomial fit. We also discuss the approach to tuning the MEMCPU Platform for problems beyond the reach of today's most advanced methods. Finally, basic analysis of the acceleration expected from an ASIC implementation is provided and suggests the possibility of real time factorization of large biprimes.

\end{abstract}

\maketitle


\section{Introduction} \label{sec:intro}

Electronic computing machines have played a crucial role in shaping the contemporary technological landscape \cite{eckert_electronic_1947, goldstine_electronic_1946}, and for almost a century, the von Neumann architecture \cite{von_neumann_first_1993} has served as the standard reference. Introduced in the mid-20th century, this architecture comprises essential components, including input and output modules, a central processing unit, and a memory bank. It has facilitated the development of general-purpose computing machines, where sets of instructions (programs) are stored in memory and executed by the CPU, with data continuously exchanged between memory and the processing unit \cite{hennessy_computer_2017}. Although the von Neumann architecture has proven versatile \cite{arora_computational_2009}, it is not without its inherent limitations, most notably the well-known "von Neumann bottleneck" \cite{hennessy_computer_2017, backus_can_1978}, which constrains system throughput due to data transfer between the CPU and memory, thereby resulting in energy inefficiency \cite{hennessy_computer_2017, hennessy_new_2019, horowitz_11_2014}. Additionally, as certain computational problems increase in size, the efficiency of von Neumann architectures becomes limited \cite{arora_computational_2009}, rendering certain problems intractable. Consequently, the demand for enhanced performance, energy efficiency, and reliability in computing devices has spurred the investigation of novel paradigms and alternative computing solutions.

In recent times, both academia and industry have put forth diverse innovative methodologies to overcome the limitations of traditional von Neumann architectures and meet the escalating demands for computational capabilities. These solutions encompass specialized hardware and architectures tailored for specific applications, such as graphical processing units (GPUs) \cite{nickolls_gpu_2010, singh_survey_2014} that were initially devised for graphical tasks but are now extensively employed in various domains, including simulations and machine learning \cite{lecun_deep_2015, goodfellow_deep_2016, sze_hardware_2017, sze_efficient_2017}. Moreover, there has been a revived interest in neuromorphic computing, where asynchronous digital or analog circuits \cite{yu_neuro-inspired_2018, burr_neuromorphic_2016} emulate brain neural networks \cite{furber_spinnaker_2014, davies_loihi_2018, boybat_neuromorphic_2018, tavanaei_deep_2019}. In the pursuit of mitigating the von Neumann bottleneck, the concept of near-memory computing \cite{singh_near-memory_2019} has emerged as a promising approach, closely related to in-memory computing \cite{ielmini_-memory_2018, traversa_dynamic_2014, pershin_memcomputing_2015, sebastian_memory_2020, le_gallo_64-core_2023, caravelli_global_2021}. However, pushing the boundaries even further, the notion of memcomputing has been introduced as a non-Turing paradigm \cite{traversa_universal_2015}, representing the most idealized model of computational memory \cite{traversa_universal_2015, di_ventra_perspective_2018, pei_universality_2019}.

The diverse array of emerging computing approaches presents captivating possibilities for meeting the impending computational demands, albeit lacking a definitive front runner at this stage. The exploration of alternative computing paradigms, such as \textit{p}-bits \cite{camsari_stochastic_2017} and coupled oscillators \cite{goto_new_1954,goto_parametron_1959,von_neumann_non-linear_1954,wigington_new_1959}, has exhibited promise, while several specialized hardware solutions are actively under development \cite{csaba_computing_2018, wang_oscillator-based_2017, chou_analog_2019, mallick_using_2020, csaba_coupled_2020, chen_dynamically_2002, shukla_pairwise_2014, mohseni_ising_2022, goto_combinatorial_2019}. As the exponential growth in computing power demand continues, it becomes increasingly apparent that the evolution of computing devices and architectures will exert a pivotal influence on the future technological landscape. This manuscript delves into the critical role that memcomputing can play in surmounting these challenges, offering wide-ranging industrial applications.

While ordinary logic gates constitute the foundation of the present electronic industry, which has made remarkable advancements, current architectures still face exponential time constraints when tackling NP-hard optimization problems. In contrast, memcomputing based solutions have shown potential on mitigating the exponential complexity of these problems \cite{traversa_evidence_2018, sheldon_stress-testing_2020, traversa_aircraft_2019, sheldon_taming_2019, manukian_accelerating_2019, bearden_efficient_2020}. Memcomputing can be realized in analog \cite{traversa_memcomputing_2015} or digital form \cite{traversa_polynomial-time_2017}. Its most effective embodiment is represented by self-organizing gates (SOGs) \cite{traversa_polynomial-time_2017, traversa_memcomputing_2018, manukian_memcomputing_2017}. These gates are engineered to achieve equilibrium based on prescribed relations at their terminals \cite{traversa_polynomial-time_2017, traversa_memcomputing_2018}. Networking SOGs creates self-organizing circuits (SOCs) \cite{di_ventra_perspective_2018, traversa_polynomial-time_2017, traversa_memcomputing_2018} adept at solving complex combinatorial optimization problems \cite{traversa_evidence_2018, sheldon_stress-testing_2020, traversa_aircraft_2019, sheldon_taming_2019, manukian_accelerating_2019, traversa_aircraft_2019, traversa_oil_2020, traversa_drone_nodate, traversa_aircraft_2023}. Remarkably, these gates can be realized in silicon emplying the same electronic components as ordinary electronic circuits. We formulate our problems as Integer Linear Programming (ILP) equations, which can be directly mapped to a Self-Organizing circuit. Our MEMCPU\texttrademark{} Platform \cite{noauthor_memcomputing_2023} automates all of the aforementioned processes.

The MEMCPU Platform has exhibited remarkable success in solving challenging logistics problems pertinent to the oil \& gas industry \cite{traversa_oil_2020}, aircraft scheduling \cite{traversa_aircraft_2023, traversa_optimizing_2021}, and swarm optimization for drones \cite{traversa_drone_nodate}. These are examples of combinatorial optimization problems encountered in industry that are often intractable for state-of-the-art solvers \cite{noauthor_gurobi_2023, noauthor_mathematical_2023}. This renders them impractical for industry partners aiming to apply them at a larger scale. In contrast, our MEMCPU Platform demonstrates a polynomial-like scaling for most of these problems, resulting in substantial savings in labor and allocated resources for these specific tasks \cite{traversa_oil_2020, traversa_aircraft_2023, traversa_optimizing_2021, traversa_drone_nodate}.

Among intractable combinatorial problems, large-scale prime factorization is a well-known challenge, the study of which is critical to guaranteeing secure encryption systems in numerous technologies. Any algorithm capable of factoring large biprimes (products of two prime numbers) within a reasonable time frame poses a significant threat to several current encryption methods, particularly those relying on RSA-based encryption \cite{milanov_rsa_2009, technology_digital_2023}. Quantum Computers \cite{benioff_computer_1980, feynman_simulating_1982, manin_computable_1980}, leveraging Shor's algorithm \cite{shor_algorithms_1994}, are considered a potential risk to RSA encryption \cite{arora_computational_2009}. There have been some approaches based on variational algorithms \cite{anschuetz_variational_2019}, however, the realization of a fault-tolerant and scalable quantum computer remains a distant aspiration \cite{dyakonov_when_2019}, contingent on considerable experimental breakthroughs. Presently, sieve methods represent the state-of-the-art algorithms showing promise, with the general number field sieve method \cite{lenstra_development_1993} being the most effective. Nevertheless, even these methods struggle to factor a 2048-bit RSA key within a sensible timeframe \cite{kleinjung_factorization_2010, zimmermann_factorization_2020}, and past instances have taken almost 2700-CPU-years to factor an 829-bit number using computer clusters \cite{zimmermann_factorization_2020}.

Our approach converts prime factorization into an ILP problem which is then conveniently mapped into a network of SOGs, which are the core of the MEMCPU Platform. This is prepared according to problem structures by tuning SOG design parameters. 
Finally, the circuit dynamics are initialized for a given input problem and run until MEMCPU converges to an equilibrium and hence the ILP solution is reached. If the structure of the problem changes with size, as for the factorization, the design parameter tuning process is repeated at each size. Details are discussed throughout this manuscript. 

In \cite{rocutto_assessing_2023} preliminary results of the MEMCPU Platform were reported. In that case, no size-dependent design parameter tuning was performed and no different structures of the ILP problem were considered during the preparation phase, and an older and less performant design parameter tuner was employed. 
By employing this whole and more advanced approach, the MEMCPU Platform is now seen to converge more efficiently to solutions of instances of factorization problems using a similar ILP formulation. The time-to-solution is observed to grow slowly with problem size. Empirically, the scaling follows a low-degree polynomial, a 5th-order polynomial fitting, while an example state-of-the-art ILP solver exhibits scaling following a 20th-order polynomial. The MEMCPU hardware, as estimated from the simulation results, appears to follow the scaling of a 3rd-degree polynomial, which would correspond to breaking 2048-bit RSA in just a few minutes. However this analysis was performed only up to 60-bit biprimes mainly due to budget constraints. 

On the other hand, leveraging congruence methods for factorization further enhances the MEMCPU Platform results. We could push our analysis up to 300 bit biprimes before running into budget constraints. The scaling trend fitted by a 2nd-degree polynomial in this case, suggests reaching RSA-relevant sizes at the time scale of weeks of simulation or real time in hardware implementation. Although these initial results show great promise, constructing and scaling to higher numbers of bits present their own challenges, as discussed in this article. However, as far as our understanding goes at this point, we do not see any physical or mathematical roadblock that would prevent continuing this trend to larger sizes.

The outline of the manuscript is as follows: In section \ref{sec:mem_tech}, we introduce the MemComputing technology. In section \ref{sec:primefactorization}, we describe the problem, our formulation, and the timing results. In section \ref{sec:circuit_design} we describe effective methods of tuning. 

\section{MemComputing Technology} \label{sec:mem_tech}

\subsection{Brief Overview of Self-Organizing Gates}\label{sec:level2}

The MEMCPU\texttrademark{} Platform finds solutions to problems cast in Integer Linear Programming (ILP) format. However, before discussing the ultimate version of the technology used for this project to handle ILPs, let us briefly introduce a simplified version of the MEMCPU Platform for solving logic functions. This platform emulates networked Self-Organizing Logic Gates (SOLGs) \cite{traversa_polynomial-time_2017, ventra_self-organizing_2018}. Each SOLG is an electronic circuit whose voltages at the terminals encode variables of the problem, and the SOLG drives these voltages to satisfy a Boolean relation among the variables\cite{traversa_polynomial-time_2017, ventra_self-organizing_2018}. 

\begin{figure}

\subfloat[\label{fig:structure_gate_a}]{%
  \includegraphics[width=0.3\columnwidth]{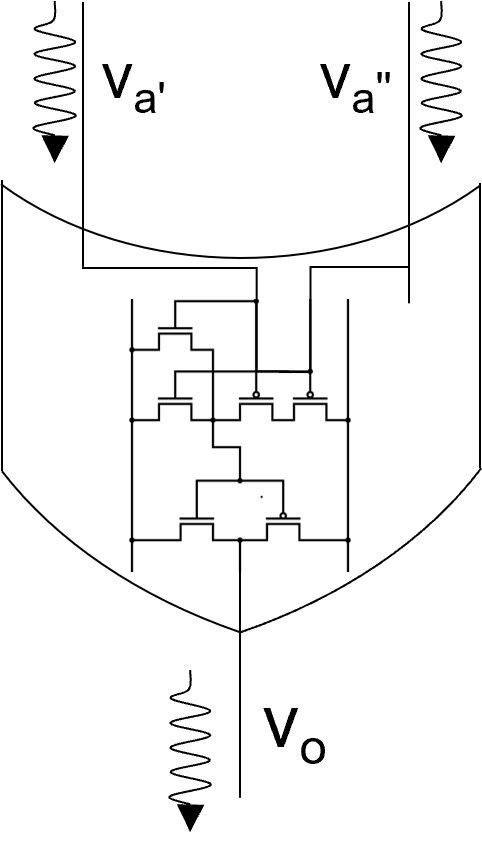}%
}
\hspace*{33pt}
\subfloat[\label{fig:structure_gate_b}]{%
  \includegraphics[width=0.3\columnwidth]{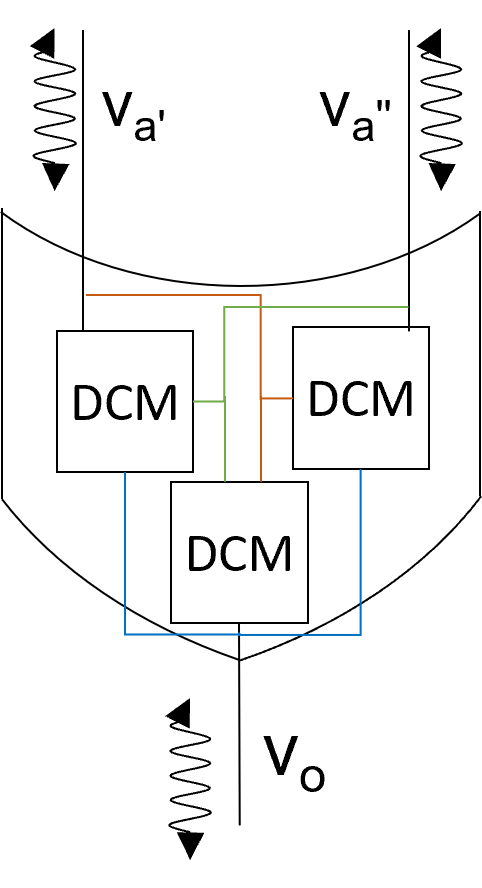}%
}
\caption{The structures of a standard logic gate (a) and a self-organizing logic gate (SOLG) (b). 
(a) A standard logic gate composed of transistors that sets the voltage of the output terminal depending on the input terminals. The wavy arrows indicate the flow of meaningful signals through the gate.
(b) A SOLG with a Dynamic Correction Module (DCM) connected to each terminal that generates feedback according to the signals it receives from the other terminals. The feedback ceases when the gate’s terminals satisfy the logic relation, leading to the term “self-organizing”.
The building blocks of both gates are standard electronic devices and both encode information through thresholds. SOLGs, however, support asynchronous operation, superposition of input-output signals, and cooperative parallel computation.
 }
\label{fig:structure_gate}
\end{figure}

\begin{figure}[b]
\includegraphics[width=0.33\textwidth]{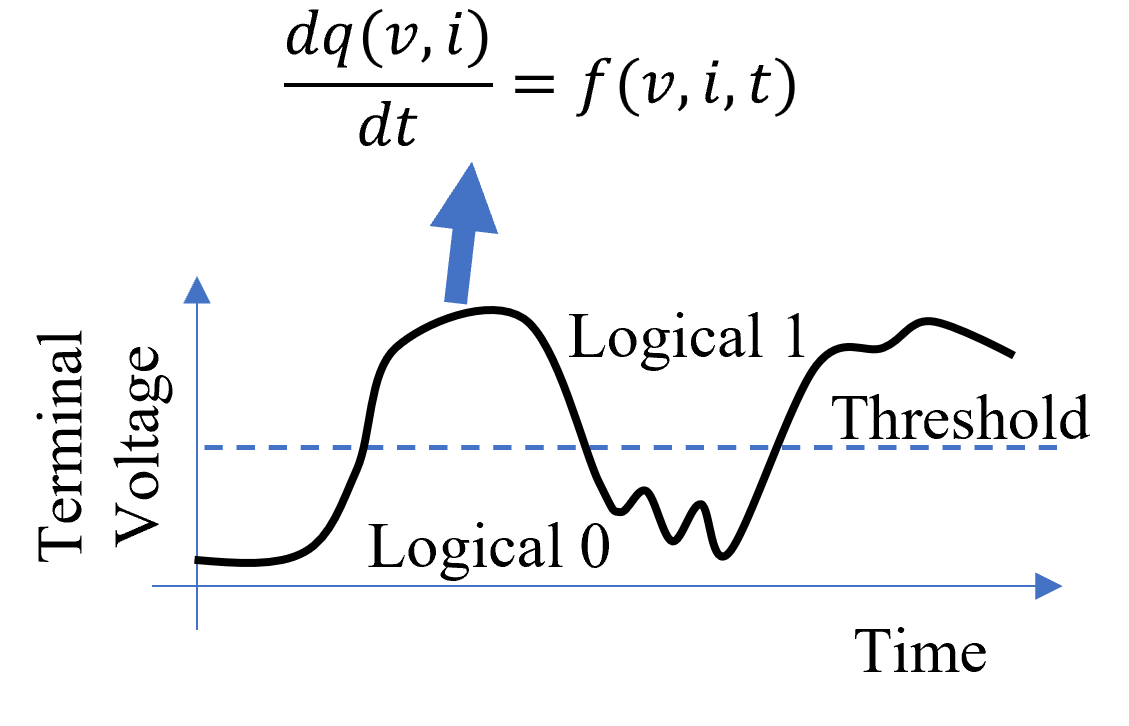} 
\caption{Information encoding through thresholds. As in a standard logic gate, SOLGs encode information through thresholds, i.e., when a terminal voltage is above a defined threshold, it is in the logical state 1, otherwise in the logical state 0. The full voltage dynamics is determined by a system of differential equations among voltages and currents as in any electronic circuit.}
\label{fig:threshold}
\end{figure}

To give a better idea, let us first consider a standard logic gate as in Figure \ref{fig:structure_gate_a}. This is a traditional electronic circuit composed of transistors that sets the voltage $v_{o}$ of the output terminal depending on how input terminals $v_{a’}$ and $v_{a''}$ are set. It does not work in reverse, which means that we cannot set the voltage at $v_{o}$ and expect to find something meaningful at $v_{a’}$ and $v_{a''}$. The wavy arrows in Fig. \ref{fig:structure_gate_a} indicate the flow of meaningful signals through the standard logic gate. 

SOLGs (Figure \ref{fig:structure_gate_b}) have similarities and major differences with standard gates. The most prominent similarities are:
\begin{enumerate}
    \item The building blocks of both gates are standard electronic devices (transistors, resistors, etc.).
    \item Both gates encode information through thresholds: if the terminal voltage is above a threshold, it encodes a logical 1 otherwise it encodes a logical 0 (Figure \ref{fig:threshold}).
\end{enumerate}

These are important shared traits. The first implies that the emulation can be performed quite efficiently using standard modern computing hardware. The second shows that we can use the MEMCPU Platform to encode problems in a digital form. This avoids well-known precision issues associated with analog computers. 

For clarity, what is meant by an efficient emulation should be explained. In this case, “efficient” means that it has a polynomial overhead with respect to the actual physical system, i.e., the circuit as realized in silicon.  
In contrast, for example, simulating a fault-tolerant quantum computer requires exponential overhead on modern hardware, which makes large scale simulations impractical. Given the polynomial overhead, the silicon version of the MEMCPU Circuit is expected to be several orders of magnitude faster than the MEMCPU Emulation that we have today. And, considering that an SOG circuit does not require access to an external memory bank during computation (the computation is within the SOG network), the energy reduction would be significant, since this eliminates the von Neumann bottleneck. The result would be an ultra-high performance and ultra-low power computational device.

\begin{figure*}
\includegraphics[width=0.75\textwidth]{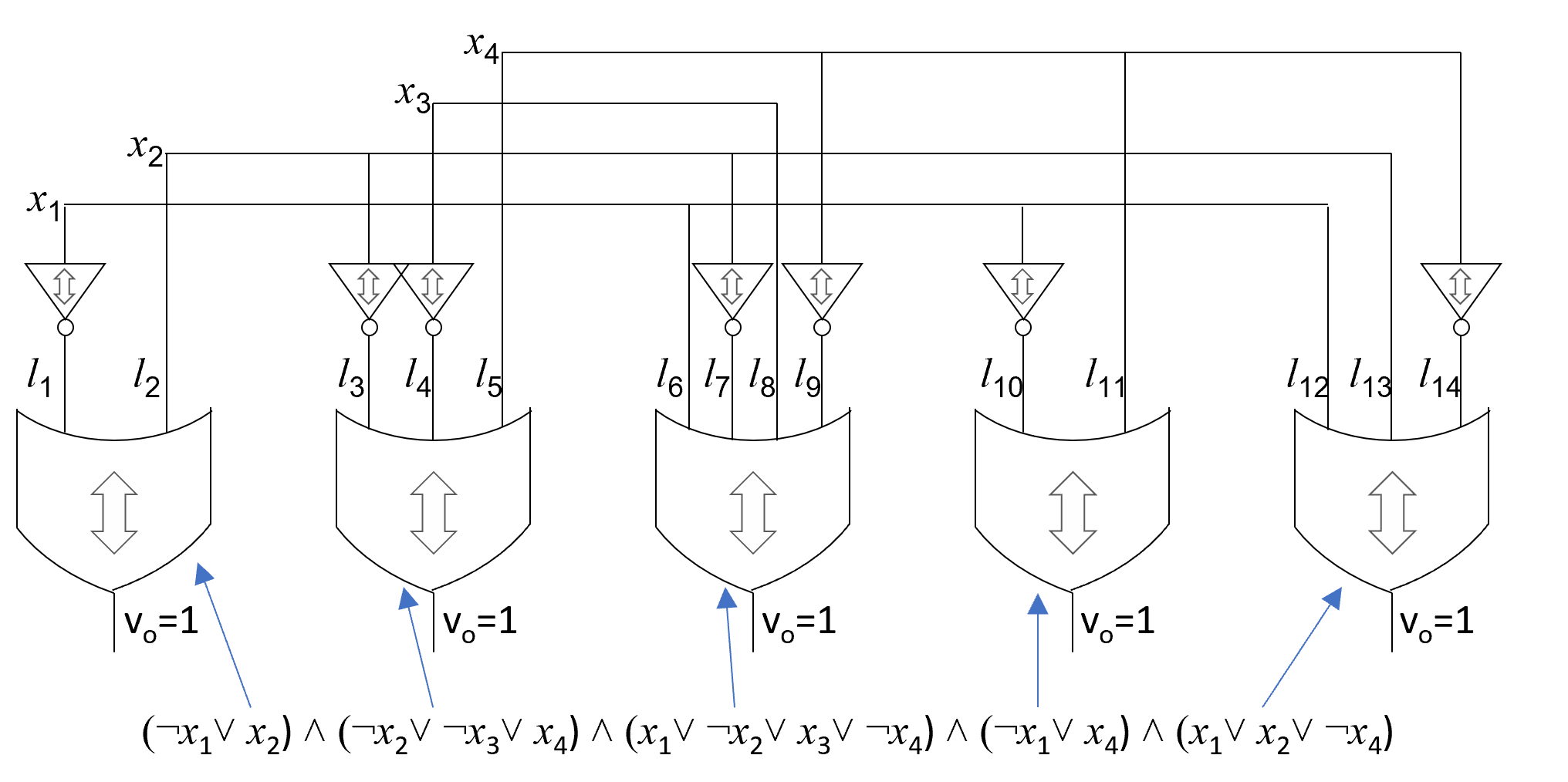} 
\caption{ A logic formula represented in conjunctive normal form (CNF) using Self-Organizing (SO) NOT and multiterminal OR gates. By setting the output of the SO multiterminal ORs to 1, the circuit finds, if it exists, the assignments at the SO OR gate terminals that satisfy all gates simultaneously and therefore the CNF formula. The SOLGs work cooperatively to find the assignment that satisfies all SOLGs, exchanging information about their state at each instant and adapting accordingly. This cooperation is achieved through superpositions of all signals arriving from the surrounding gates while simultaneously sending signals about their state, allowing for pure asynchronous communication.}
\label{fig:logic_formula_cnf}
3\end{figure*}

Significant differences between SOLGs and standard gates are:
\begin{enumerate}
    \item SOLGs are asynchronous and do not require a clock.
    \item SOLGs support the superposition of input-output signals at each terminal. Standard gates have dedicated input and output terminals.  
    \item SOLGs compute cooperatively in parallel with other SOLGs, while standard gates compute sequentially. 
\end{enumerate}

These differences are profound and define the way SOLGs work. Figure \ref{fig:logic_formula_cnf} illustrates how these features impact the working principle of SOLGs. 

Consider the logic formula represented in Conjunctive Normal Form (CNF) in Fig. \ref{fig:logic_formula_cnf}. Each variable in the formula is a binary variable, and the goal is to find an assignment that makes the whole formula true. To find this assignment, we can use Self-Organizing (SO) NOT and multi-terminal OR gates connected as depicted in Fig. \ref{fig:logic_formula_cnf}.  By setting the output of the SO multi-terminal ORs to 1, we require that the circuit finds the assignments at the SO OR gate terminals that satisfy all gates simultaneously and therefore the CNF formula \cite{ventra_memcomputing_2022}.   

This example helps us understand the implications of the differences listed above. First, we notice that the SOLGs in Fig. \ref{fig:logic_formula_cnf} need to work cooperatively to find the assignment that satisfies all SOLGs. (Note that standard logic gates certainly could not be used in this setting, due to the differences listed above.) For a circuit like the one in Fig. \ref{fig:logic_formula_cnf} to work properly, i.e., to have gates cooperate effectively, the gates should exchange information about their state at each instant so that they can adapt accordingly. This cooperation can be obtained only if each terminal of the gates receives signals carrying that information from the other gates. When designed in that way, the system can have the best response, because it has continuous knowledge of the rest of the circuit. That is, each gate dynamics is affected by changes to the surrounding gates as they occur.

To achieve the collaborative process just described, the terminals of the SOLGs must therefore support superpositions of all signals arriving from the surrounding gates while simultaneously sending signals about their state. This means that the SOLG terminals support superposition of both input and output signals (difference \#2). The fact that each SOLG needs continuous knowledge of the state and internal dynamics of all other surrounding SOLGs means that the communication is continuous in time, and therefore it cannot be clocked. This ultimately implies pure asynchronous communication (difference \#1). 

As depicted here, the working principles of SOLGs are achieved using feedback loops. Each SOLG has what we call Dynamic Correction Modules (DCMs), one connected to each terminal, that generate feedback according to the signal it receives from the other terminals of the gate (see Fig. \ref{fig:structure_gate_b}). We use the term “self-organizing” to describe this activity: DCMs drive voltages autonomously via feedback loops, and once the SOLG satisfies the logical relation, the feedback ceases. Finally, where the terminal of one SOLG connects to one or more other SOLG terminals, the feedback from the DCMs of each terminal is propagated in the entire circuit, so the circuit behaves as a cohesive network to satisfy all SOLGs at once and minimize the global feedback \cite{traversa_polynomial-time_2017, ventra_self-organizing_2018}. However, certain energy balance and signal propagation properties must be satisfied for the circuit to work properly \cite{di_ventra_topological_2017, sheldon_taming_2019}. This requires finding design parameters that guarantee such conditions. Aspects of this are discussed in Section \ref{sec:circuit_design}. (More information for particular applications is available in case studies, whitepapers, and peer-reviewed articles here \cite{noauthor_resources_2020}). 

\begin{figure*}
\includegraphics[width=0.55\textwidth]{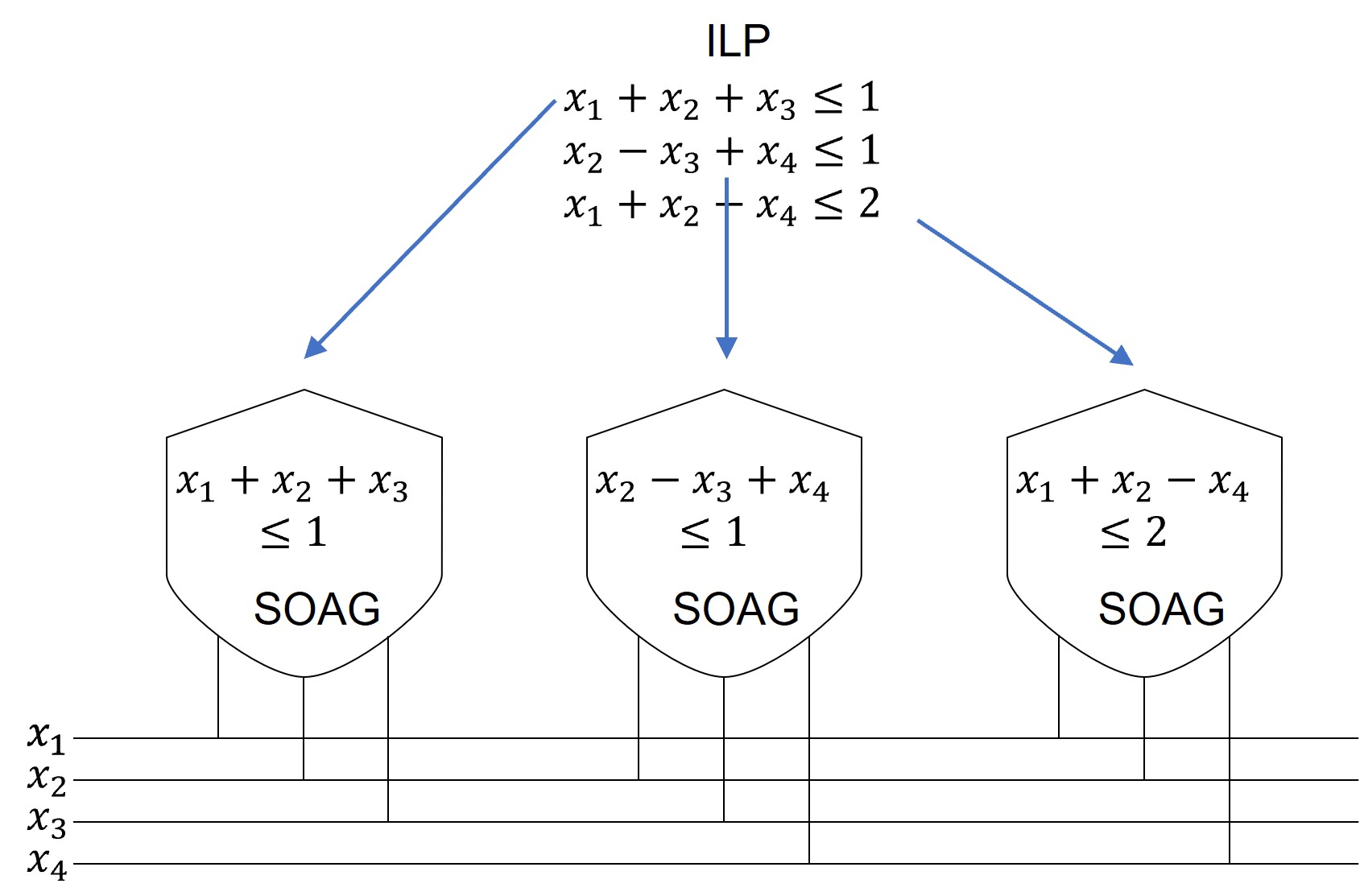} 
\caption{A circuit assembled using Self-Organizing Algebraic Gates (SOAGs) designed to solve linear inequalities. Like SOLGs, SOAGs have DCMs that create feedback on the terminal they are connected to based on signals from the other terminals and turn off when the gate is satisfied. SOAGs are satisfied when the state of their terminals satisfies a linear inequality, allowing them to be used to solve problems cast in the form of Integer Linear Programming (ILP), a versatile format for combinatorial optimization problems widely used in industry.}
\label{fig:soag_circuit}
\end{figure*}

Finally, we introduce Self-Organizing Algebraic Gates (SOAGs) \cite{traversa_memcomputing_2018} designed to solve linear inequalities. Like the SOLGs, SOAGs have DCMs that create feedback on the terminal they are connected to based on signals from the other terminals, and they turn off if the gate is satisfied. The main difference is that an SOAG is satisfied when the state of its terminals (they still encode binary variables like SOLGs) satisfies a linear inequality. Using these gates, we can assemble circuits like the one in Figure \ref{fig:soag_circuit} and use them to solve problems cast in the form of Integer Linear Programming (ILP), a much more versatile format for combinatorial optimization problems widely used in industry.

\subsection{Other applications}

We have applied the MEMCPU Platform to solve many combinatorial optimization problems relevant to academia \cite{traversa_evidence_2018, sheldon_taming_2019} as well as industry \cite{traversa_aircraft_2019, traversa_optimizing_2021, traversa_aircraft_2023, traversa_oil_2020}, and military \cite{traversa_proliferated_2022, traversa_drone_nodate}. We focus on problems that are usually hard, if not intractable, for state-of-the-art ILP solvers like IBM CPLEX \cite{noauthor_mathematical_2023} and Gurobi \cite{noauthor_gurobi_2023}. Our approach usually outperforms those solvers for problems where it is hard to apply shortcuts or heuristics to simplify the problem (both CPLEX and Gurobi rely heavily on those methods to accelerate their solution when possible). Those solvers commonly rely on branch and cut, the “smartest” brute force method to solve ILPs. The compute time required and memory consumption can grow exponentially for these standard methods~\cite{klotz_practical_2013, basu_complexity_2022}. We have proven the efficiency of the MEMCPU Platform on many scenarios for a variety of customers \cite{noauthor_resources_2020}. We summarize three of them in Appendix \ref{app:industrial_application} to give an idea of the potential of our approach.

\section{Prime Factorization} \label{sec:primefactorization}

RSA cryptography is a public-key cryptosystem that is widely used for secure data transmission. It was invented by Ron Rivest, Adi Shamir, and Leonard Adleman in 1971. The security of RSA relies on the practical difficulty of factoring the product of two large prime numbers, the “factoring problem” \cite{milanov_rsa_2009}. The RSA algorithm raises a message to an exponent $e$, modulo a composite number $n$, whose factors are not known. Thus, the task can be described as finding the $e-th$ roots of an arbitrary number, modulo $n$ \cite{milanov_rsa_2009}. For large RSA key sizes (\emph{e.g.} 1024 bits or more), no practical method for solving this problem is known; if an efficient method is ever developed, it would threaten the current or eventual security of RSA-based cryptosystems—both for public-key encryption and digital signatures \cite{milanov_rsa_2009, technology_digital_2023}. The most efficient method known to solve the RSA problem is by first factoring the modulus $n$ \cite{milanov_rsa_2009, technology_digital_2023}.

\subsection{Existing Approaches}

It is estimated that with current technology using the best-known algorithm (general number field sieve, GNFS), factoring a 2048-bit RSA key would take longer than the age of the universe. For reference, a 768-bit RSA key was factored by researchers in 2009, and that required more than two years of computation using hundreds of multicore CPUs \cite{kleinjung_factorization_2010}. After a decade, a new record was set in 2020 \cite{zimmermann_factorization_2020} where factoring an 829-bit number (RSA250) required 2700 CPU-years (it was factored parallelizing the GNFS over CPUs of supercomputer centers in France, Germany and California). Over more than a decade, we have only seen a very small improvement (i.e., 61 bits). This is mainly due to the lack of innovative new algorithms, and the fact that conventional computer performance is reaching a plateau. Factoring a 1024-bit number is considered to be potentially within reach of the resources of a large nation-state within the next decade using this method. However, factoring a 2048-bit number remains well beyond the capability of any known existing technology in the foreseeable future \cite{kleinjung_factorization_2010}. 

On the other hand, a hypothetical fault-tolerant quantum computer, if able to implement Shor’s algorithm, could factor a 2048-bit RSA key in several hours or less \cite{shor_algorithms_1994}. Shor’s algorithm is the best-known quantum algorithm for factoring large numbers and is exponentially faster than the most prominent classical algorithms. Shor’s algorithm could run in polynomial time (specifically, time $\mathcal{O}((\log n)^3 (\log \log n) (\log \log \log n)) $, for a number $n$ \cite{beckman_efficient_1996}). In contrast, the best-known classical factoring algorithm, the GNFS, runs in super-polynomial time complexity $\mathcal{O}(\exp( 1.9 (\log n)^{1/3} (\log \log n)^{2/3}) )$ \cite{lenstra_development_1993}. A fully fault-tolerant quantum computer is a significant technological challenge, and in 2021 industry consensus was that it was anywhere from 10 to 30 years in the future, and, even then it is still questioned whether it will break RSA 2048 encryption \cite{mosca_2022_2022}.

\subsection{MemComputing Approach} \label{subsec:mem_approach}

MemComputing got limited funding for this project as part of a Phase II SBIR with an Air Force Intelligence Group in which we developed a MemComputing-based solution to perform prime factorization. It was built upon the cloud-based MEMCPU Platform. The MEMCPU Platform is programmed by submitting a mathematical model formulated using Integer Linear Programming (ILP) combined with specialized design and control parameters for the MemComputing Circuit. Part of the effort of properly programming the MEMCPU Platform is spent testing and profiling the mathematical model, identifying inefficiencies, and then iterating through improvements to reach the best fit. Once the mathematical model has been established, circuit design and control parameters need to be optimized to produce convergence as quickly and robustly as possible. Iterative testing and profiling is performed using advanced optimization techniques to improve the circuit’s performance with knowledge of the mathematical formulation. Following this process, by the end of Phase II, the MemComputing solution was addressing up to 300-bit factorization problems.

\subsubsection{Direct model: description and performances}\label{subsubsec:direct_model}

We began with a direct approach to the factorization problem. This involved formulating solutions to the equation $pq=n$ where $p$, $q$ are unknown integers and $n$ is the known integer. There are many ways to transform this equation into an ILP model. They all involve binarization of the integers and then different ways to express the binary products and composing the equations. For example, the binary products $p_j q_k=s_{jk}$  can be expressed using 2 inequalities like, 
\begin{align}\label{eq:two_inequal}
   p_j+q_k \leq s_{jk}+1, \quad   p_j+q_k \geq 2s_{jk} .
\end{align}
However, it is easy to see that the first inequality can be substituted by $p_j+q_k \leq 2s_{jk}+1$ providing the same solutions. This is only an example of many different variations we can use to express the binary product. We tested many variations and eq. \ref{eq:two_inequal} worked better with the MEMCPU Platform \cite{[{Detailed results can be found in the project reports. Limited availability upon request.}  ] noauthor_contact_nodate}.

Considering the expansion $p=\sum_{j=0}^{N_p-1} 2^j p_j$, and similar ones for $q$ and $n$ and the binary products $s_{jk}$, we can express the product $pq=n$ through a set of equalities summing up all binary products belonging to groups of expansion coefficients. The general equations read
\begin{align}\label{eq:prime_product}
      \sum_{g=MG}^{(M+1)G-1} 2^{g-MG} \left(\sum_{j+k=g} s_{jk} + \sum_{m=0}^{M-1} r_{mg}\right) \nonumber \\  
      =  \sum_{g=MG}^{(M+1)G-1} 2^{g-MG} n_g + \sum_{g=G}^{G+\lfloor \log_2 \sup(\text{lhs}) \rfloor +1} 2^g r_{Mg},  
\end{align}
where $G$ is the size of the coefficient group, $M$ is the $M$-th group of coefficients, $r_{mg}$ are the binary coefficients of the remainders of the additions, and $\sup(\text{lhs})$ is the upper bound that the left-hand side of eq. \ref{eq:prime_product} can reach.

Also, in this case eq. \ref{eq:prime_product} is not the only way to express the product $pq=n$ as an ILP model. We have tested many other variants, including more naive binary product representation using 3 bits adders or more involved variants using Karatsuba or Montgomery products. However, our tests showed that representation \ref{eq:prime_product} with $G=2$ was the best choice for the MEMCPU Platform \cite{[{Detailed results can be found in the project reports. Limited availability upon request.}  ] noauthor_contact_nodate}.

\subsubsection*{Benchmark and Scaling results}

\begin{figure*}
     \centering
\subfloat[\label{fig:scaling_a}]{%
  \includegraphics[width=.99\columnwidth]{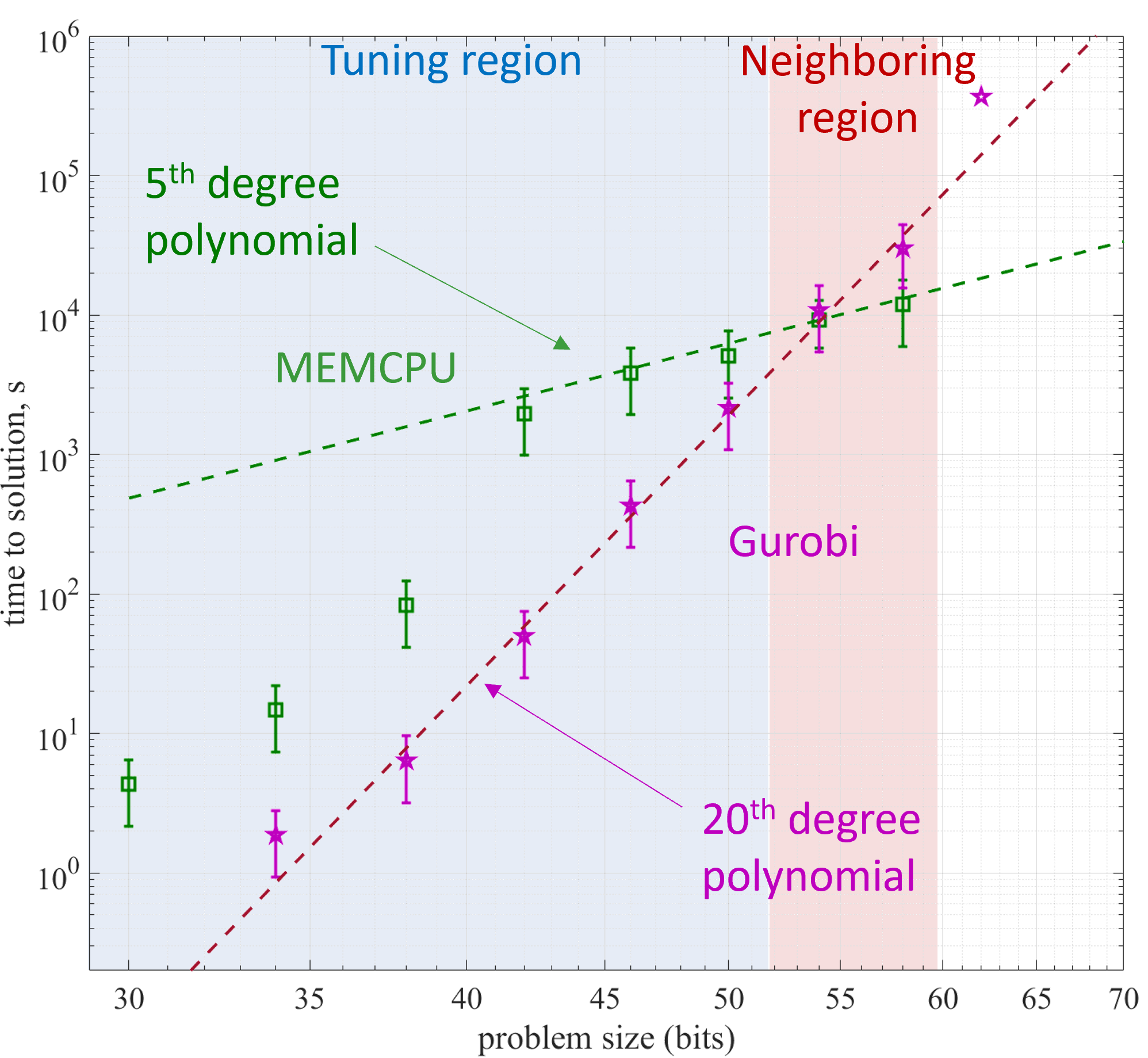}%
}
\hfill
\subfloat[\label{fig:scaling_b}]{%
  \includegraphics[width=1.01\columnwidth]{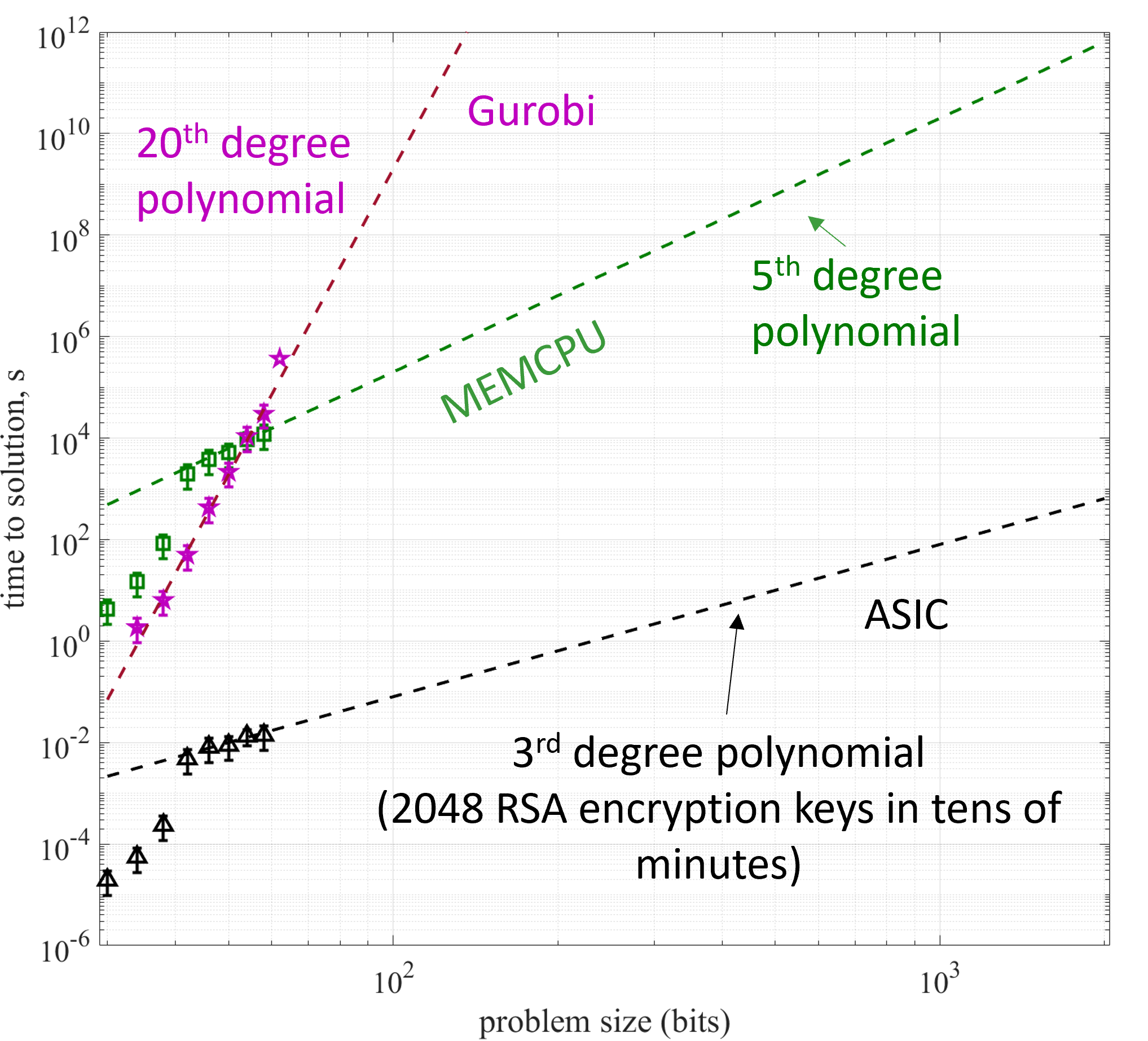}%
}
\caption{Comparison of the scaling results for factorization problems between the MEMCPU Platform and the Gurobi solver. (a) Raw timings (symbols) and polynomial fits (dashed lines). The Tuning region is where the design parameters were fine-tuned for each size, while the Neighboring region employed the design parameters of smaller problem sizes, which, however, maintained about the same performance. (b) Same data, with extrapolation of the timing trend to 2048 bits, and the forecasted ASIC timings based on the MEMCPU Platform results. }
\label{fig:scaling}
\end{figure*}

The benchmarks we generated were designed to be among the hardest factorization problems. We used prescriptions to generate primes for strong RSA encryption \cite{technology_digital_2023}. The procedure involved generating a random string of bits and adding a 1 to both the beginning and end of the string to ensure that the number was odd and had a given length in bits. Then, we performed a primality test to select only prime random numbers. As a result, this selection would provide primes with a uniform distribution of bits. We then selected pairs of primes with the same length, whose product would produce an integer with a length exactly double that of the two primes.

In Figure \ref{fig:scaling}, we report the timing results of this formulation. As the model is ILP, we also ran the same benchmark using the Gurobi ILP solver \cite{noauthor_gurobi_2023}. As expected, the performance from Gurobi quickly diverged and, in a log-log plot (time to solution \emph{vs.} length of the product in bits), it was fit by a polynomial of degree 20. 

On the other hand, the MEMCPU Platform displays two different scaling trends: one at a lower number of bits and the other at a higher number. This difference in scaling trends results from the ability of the preprocessing to simplify smaller problems more effectively, making them equivalent to even smaller ones. Indeed, the MEMCPU Platform includes a basic routine that simplifies, when possible, the ILP model submitted. For the factorization model, this routine is very effective up to about 40 bits, then it loses its efficiency. On the other hand, Gurobi has a much more advanced preprocessing routine that, looking at the log files, consistently works at any size we tested, providing a smoother scaling. This effect is further amplified by the fact that we were able to find better circuit design parameters for smaller problems, given the cloud resource constraints during this SBIR and the need for further improvements to our in-house CAD toolbox (see section \ref{sec:circuit_design} for details on the design procedure). The scaling reaches a more stable behavior above about 42 bits in length, fitted by a 5-degree polynomial. 

As described in Section \ref{sec:circuit_design}, the MEMCPU Platform is tuned for each size of factorization problem since the structure of the problem varies with the size. This procedure can be quite costly and is still being improved to increase its efficiency. We tuned for all sizes in the tuning region highlighted in figure \ref{fig:scaling}a. However, due to limited resources, we performed only incomplete tuning starting from the previous size using continuation methods combined with parallel tempering (see section Section \ref{sec:circuit_design} for details) up to 50 bits. Testing further (Neighboring region in figure \ref{fig:scaling_a}), up to about 60 bits, we were able to effectively use the the design found at lower bits and still keep the same scaling properties as reported in Figure \ref{fig:scaling_a}. These results represent already a major improvement, in terms of absolute timing (orders of magnitude faster) and scaling, with respect to the ones presented in \cite{rocutto_assessing_2023} for motivations mentioned in the introduction. This proves that our approach is still prone to major improvements. For example, a more thorough parameter tuning and improvements to our in house CAD software will reduce much further the absolute timing and might even reduce the degree of the scaling. Also, tuning problems at larger sizes will extend the scaling to larger problem sizes. In addition, upgrades to the SOG design could also bring additional improvements. 

In Figure \ref{fig:scaling_b}, the same scaling is extrapolated up to 2048 bits, assuming continued tuning. The timing for the ASIC realization of the MEMCPU Platform is also reported. The ASIC timing can be easily estimated since the MEMCPU Platform, being a circuit emulator, returns the full dynamics of the circuit, including the simulated runtime. It is worth noting that, at this point in our R\&D, the forecast for the ASIC shows the possibility of solving a 2048-bit factorization problem in tens of minutes.

\subsubsection{Sieve model: description and performances}\label{subsubsec:sieve_model}

Subsequently, we took a new approach based on a congruence method sharing similarities with the quadratic sieve or the general number field sieve (GNFS) method. In this case, we use the MEMCPU Platform to return unique congruences, i.e., special relations among integers, that ultimately are used to factorize the large biprime. Note that using these congruences to factor is a well-known and standard method \cite{pomerance_smooth_2008} and the most effective known as of today \cite{zimmermann_factorization_2020}. A very interesting fact is that the sieve method has seen its computational complexity decrease because of smarter and smarter ways to compute congruences \cite{benjamin_tale_2009}. In fact, the basic Kraitchik’s method has complexity $O(\exp(\sqrt{2\log n \log\log n})$ \cite{benjamin_tale_2009}. But by making the search for the congruences a bit smarter we have the quadratic sieve, which has complexity $O(\exp(\sqrt{\log n \log\log n})$ \cite{benjamin_tale_2009}. Finally, employing the smartest sieve method, the GNFS, we arrive at the modern complexity of $O(\exp((64/9\log n)^{1/3} (\log\log n)^{2/3})$ \cite{benjamin_tale_2009}. Therefore, it would not be surprising if some new way to compute congruences will further decrease the complexity.    

\begin{figure*}
     \centering
\subfloat[\label{fig:scaling_sieve_a}]{%
  \includegraphics[width=0.98\columnwidth]{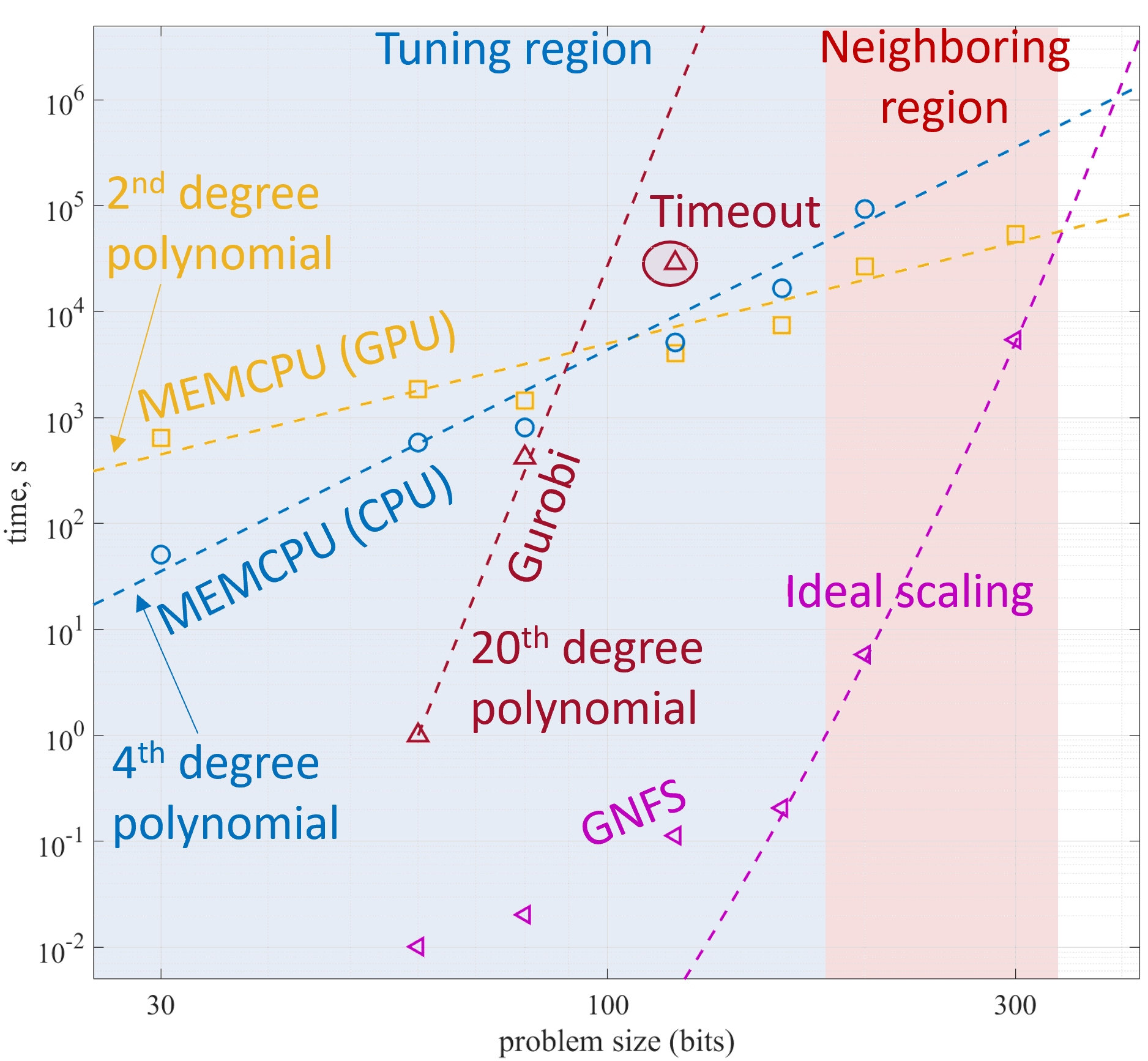}%
}
\hfill
\subfloat[\label{fig:scaling_sieve_b}]{%
  \includegraphics[width=1.02\columnwidth]{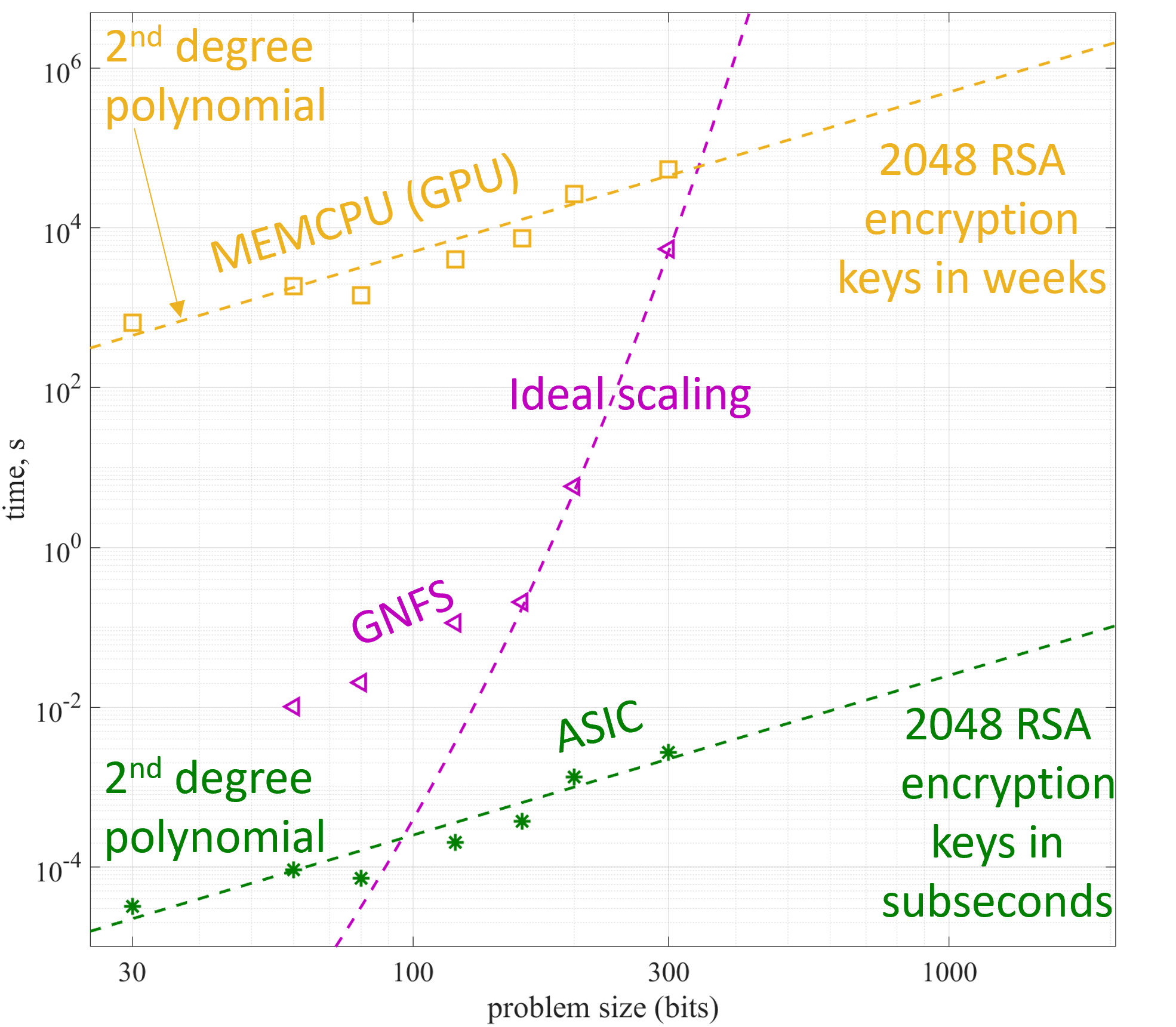}%
}
\caption{(a) Comparison of scaling results for factorization problems up to 300 bits using the MEMCPU Platform and Gurobi solver to find congruences at fixed smoothness independent of the problem size, and GNFS implemented in Msieve software. The MEMCPU Platform exhibits efficient scaling, with a 4th-degree polynomial fit for CPU emulation and a 2nd-degree polynomial fit for GPU emulation. The Tuning region is where the design parameters were fine-tuned for each size, while the Neighboring region was able to use the design parameters of smaller problem sizes, which, however, maintained about the same performance. Gurobi timed out if pushed to larger problem sizes and has a 20th-degree polynomial fit like we had with the direct model. (b) The ASIC realization of the MEMCPU Platform shows a scaling similar to the GPU emulator but the extrapolation points to solving 2048-bit factorization problems sub-second. (This plot differs somewhat from an earlier version which used an older ILP formulation.)}
\label{fig:scaling_sieve}
\end{figure*}

To use the MEMCPU Platform to find these congruences, we have developed an integer linear programming (ILP) formulation whose solutions are congruences. Therefore, it is enough to solve the ILP multiple times and then factorize the biprime using the same procedure as the standard sieve methods\cite{lenstra_development_1993, pomerance_smooth_2008}. An example of factorization using GNFS is reported in Figure \ref{fig:scaling_sieve} (a more detailed discussion is in the next subsection). 

In this case, the problem we discuss is not unique, but it is the one that worked best during this project. More tests and other models/equations for finding useful congruences can be found in the biweekly reports from the SBIR \cite{[{Detailed results can be found in the project reports. Limited availability upon request.}  ] noauthor_contact_nodate}. The basic equation for congruences we consider is similar to the one used by Schnorr \cite{schnorr_fast_2021}. Although we use a similar form for the congruences, our approach to compute them is completely different and does not rely on any of the methods mentioned in that article. However it is useful as reference for the specific form of congruence we discuss here. The congruence reads:
\begin{align}\label{eq:quad_sieve}
   x+kn= y     , 
\end{align}
where $n$ is the integer to be factorized, $k$ is a positive integer (in \cite{schnorr_fast_2021} $k$ is required to be $b-$smooth, however it is not necessary in our case) and $x$ and $y$ are required to be $b-$smooth numbers, i.e., integers whose prime factors are all smaller than or equal to $b$. We denote with $\pi(b)$ the prime counting function, i.e., the number of primes smaller than or equal to $b$. Also, for any $m\le \pi(b)$ let us assume we collected $m+1$ independent congruences containing exactly $m$ unique primes in all $x_j$ and $y_j$ with $j=1,..,m+1$. The set of these $m$ primes is called the factor base $F$. By expanding  $x_j=\prod_{p\in F}p^{\alpha_p,j}$ and $y_j=\prod_{p\in F}p^{\beta_p,j}$ in the factor base $F$, without loss of generality, we can multiply each congruence by $s_j = \prod_{p\in F}p^{\alpha_p,j \mod 2}$. The new congruences read 
\begin{align}\label{eq:quad_sieve_post}
s_jx_j+s_jk_jn & = \nonumber \\
=x'^2_j + k'_jn & = s_jy_j = y'_j 
\end{align}
Therefore we are now in a situation equivalent to the quadratic form used in the quadratic sieve. We can therefore associate to each derived congruence the vector $v_j = \{\beta_p,j+(\alpha_p,j\mod2)\}$ of the exponents of the factor base $F$. We can then efficiently select a subset of these congruences using Gaussian elimination to solve 
$\sum_j e_j v_j \mod2 = \{0_p \}$, where $e_j$ is a binary variable that is 1 if a congruence $j$ is selected and 0 otherwise. Once the selection is evaluated, we can multiply side by side all congruences selected, and their product will produce a Fermat relation of the form $Y^2-X^2 = Kn$. This can be efficiently used to find the factors of $n$ by computing the $\text{gcd}( Y-X , n)$.

Standard methods would search for smooth solutions of equation \ref{eq:quad_sieve} employing sieving methods like the quadratic sieve or GNFS. However, by converting the problem in ILP we can directly require that the $x$ and $y$ are $b-$smooth. There are several ways to do this. The most generic one is requiring that both $x$ and $y$ are products of smaller and smaller integers, for example 
\begin{align}
x = \prod_k x^{(k)} \label{eq:x_expansion}\\
y = \prod_k y^{(k)} \label{eq:y_expansion}.
\end{align}
To decide how many $x^{(k)}$ and $y^{(k)}$ we need to determine the probability that a solution is b-smooth, let us consider that, in ILP formulation, we will use their binary representation to encode the products as we discussed in section \ref{subsubsec:direct_model}. Therefore, if we implement each $x^{(k)}$ and $y^{(k)}$ using a certain number of bits, then we can determine what is the probability that a solution of the ILP version of the equation \eqref{eq:quad_sieve} is $b-$smooth. For example, if $b=2^h$ and the length of each $x^{(k)}$ and $y^{(k)}$ in bits is exactly $h$, then the solution will be 100\% $b-$smooth. If some of the $x^{(k)}$ and $y^{(k)}$ are larger than $h$, then the probability can be evaluated calling out the Dickman-de Bruijn function \cite{canfield_problem_1983}. For example, if we leave all $x^{(k)}$ and $y^{(k)}$ of length $h$ except one, let's say $x^{(1)}$, with length $rh$ for some $r\ge1$, then the probability that a solution is $b-$smooth is given by $\rho(r)$, where $\rho$ is the Dickman-de Bruijn function. For rough estimates, it can be approximated as $\rho(r)\approx r^{-r}$. 
If more than one of the $x^{(k)}$ and $y^{(k)}$ are larger than $h$, then we can use the probability of independent events, that, in this case, would be the product of all Dickman-de Bruijn functions, one for each $x^{(k)}$ or $y^{(k)}$, exceeding $h$. 

If we keep the probability high enough, we then shift the complexity to finding the solutions of the ILP problem. So, if we have an efficient solver for these kind of ILP problems, then we can solve the factorization problem efficiently. 

We chose the form of the congruence like equation \eqref{eq:quad_sieve} because it offers the possibility to be implemented in ILP in a few different ways and also allows us to make some manipulations that help in keeping the problem size compact. Even if we discuss some of these aspects in this work, it is not intended to be an exhaustive description.  

A useful first manipulation is writing eq. \eqref{eq:quad_sieve} as
\begin{align}\label{eq:quad_sieve1}
   \bar xx+kn= y     , 
\end{align}
where $\bar x$ is a fixed number. For each $\bar x$ chosen appropriately, we have a different ILP problem which can play a similar role as the multiple polynomial method for the quadratic sieve \cite{silverman_multiple_1987}, allowing, for example, for independent parallelization. However, fixing $\bar x$ facilitates running  multiple problems in parallel in a better way.

For example, let us consider $b=2^h$ and $kn<2^H$. If  $2^{H-h-r}\le\bar x\le2^{H-h-r+1}$, then $x$ can be set of length (in bits) $h+r$, and therefore, if $r$ is small enough, we will not need to further break $x$ using \eqref{eq:x_expansion} providing a non trivial simplification of the problem. It is worth noting that, if this scheme is used, $r$ and the length of $k$ need to be chosen carefully to guarantee that, at least statistically, the problem \eqref{eq:quad_sieve1} has solutions. An estimate of the number of $b-$smooth solutions of eq. \eqref{eq:quad_sieve1} as a function of $r$ and $H$ can be easily done using Dickman-de Bruijn functions.

An interesting question is how do we choose $\bar x$? A convenient way can be either a preset $b-$smooth number or a square. However, there is a better choice to speed up the computation of the congruences. If we leave $x$ and/or some of the $y^{(k)}$ in the equation \eqref{eq:y_expansion} larger than $b=2^h$, then it is probable to obtain some of the congruences with one or more factors larger than $2^h$. Typically, if there is only one large factor, the congruence can be stored and then combined with other congruence having the same large factor found by chance. This is typically employed in the standard sieve methods \cite{pomerance_smooth_2008} since finding these collisions is actually quite likely, as in the birthday paradox. However, we can do better than relying on chance. We can create problems where the large factors are included in $\bar x$ providing a nontrivial speed up. A deeper analysis of the speed up deriving from this method will be a subject of future work since here we are not leveraging it yet.

It is worthwhile discussing some details of the ILP implementation of equation \eqref{eq:y_expansion}. If there are just two terms in the \eqref{eq:y_expansion}, then the implementation  of the product would be identical to the implementation discussed in section \ref{subsubsec:direct_model}. However, if there are more than two terms, we should implement it as a hierarchy of products. For example, we can use the iterative scheme 
\begin{equation}
z^{(k+1)} = z^{(k)}y^{(k+1)}\label{eq:y_grouping}    
\end{equation}
with $z^{(0)} = y^{(0)}$. This implies $z^{(K)} = \prod_{k=0}^K y^{(k)}$ and each of the $z^{(k)}$ is a product of two integers that can be implemented as in section \ref{subsubsec:direct_model}. This is not the only way to group the products, however we found that this provided better convergence with the MEMCPU Platform. Another important aspect to notice is that the sum of all lengths in bit of the $y^{(k)}$ must exceed $H = \log_2(\sup(kn))$. The more it exceeds $H$, the more solutions the ILP has (with no impact to the probability of smoothness). However, it also makes the ILP larger. Therefore, a trade-off between the ILP size and the solutions of the problems should be found. We did not do a deeper analysis to asses the optimal sum of all lengths in bit of the $y^{(k)}$ -- it will be probably matter of future work -- however we found that $20\%$ larger than $H$ provides a good trade-off in the range of the size we tested.

Finally, there is another way to express the equation \eqref{eq:y_expansion} in ILP. In this case, we can simply implement \eqref{eq:y_expansion} as a set of equations like
\begin{align}
    \bar xx + kn =& w^{(0)}y^{(0)}\nonumber\\
       &\vdots\label{eq:quad_sieve2}\\
    \bar xx + kn =& w^{(K)}y^{(K)}\nonumber
\end{align}
coupled with the constraint $y^{(k)}\ne y^{(k')}$ for each $k\ne k'$. This condition can be achieved in ILP in many ways. 
However, tests with MEMCPU Platform have shown that, with proper MEMCPU design parameters, the SOG network naturally converges to a solution with negligible likelihood to have collisions among $y^{(k')}$ if we initialise the circuit with random initial conditions. It is worth noting that this approach might need more splits $y^{(k)}$ than \eqref{eq:y_grouping} since even two different $y^{(k)}$ can have primes in common, therefore making it not necessarily true that $\bar xx + kn = \prod_k y^{(k)}$. Based on our tests discussed in the next section, we found that if the sum of all lengths in bits of the $y^{(k)}$ is $30\%$ larger than $H$ and $ \log_2 (\sup (w^{(k)})) + \log_2 (\sup (y^{(k)})) \le H+2$ then the probability that $\bar xx + kn$ has factors other than the ones in the $y^{(k)}$ is negligible when we use the MEMCPU Platform.

\subsubsection*{Benchmark and Scaling results}

We used the same benchmark described in the section \ref{subsubsec:direct_model}. However, this time we were able to push to larger problems, reaching up to 300-bit problems. In this test, we implemented the ILP model of equation \ref{eq:quad_sieve2} and solved it to find $b-$smooth congruences with $b=2^{21}$, statistically leveraging large factors, as normally done in standard sieve methods, with 27-bit cutoff . These values were kept independent of the problem size. The time reported in Figure \ref{fig:scaling_sieve} is the time to find $\pi(b)$ independent congruences with $\pi(b)$ the prime counting function. As shown in Figure \ref{fig:scaling_sieve_a}, using a best-in-class ILP solver like Gurobi, this formulation does not reduce complexity (as expected) since it is not efficient in solving this type of ILP problem and finding congruences. In fact, Gurobi’s best fit for its scaling is still a $20^{th}$-degree polynomial (the last point timed out after 8 hours without returning any congruence), which is expected since the ILP has similar structures as in the direct model.

We also compared against GNFS implemented in the Msieve software \cite{noauthor_msieve_2023}. This implementation starts following the asymptotic GNFS sub-exponential scaling at around 150-bit factorization, as shown in Figure \ref{fig:scaling_sieve_a}. Msieve software was running on an Intel i9-11950H with 128GB RAM.   

Our MEMCPU Platform proved to be very effective in finding congruences at fixed smoothness. In fact, its scaling is seen to follow a $4^{th}$-degree polynomial for the emulation on CPUs.  We were able to achieve a much lower scaling ($2^{nd}$-degree polynomial fitting) using our emulation on GPUs, as shown in Figure \ref{fig:scaling_sieve}. The change in slope comes from the fact that the GPU implementation of the MEMCPU Platform can efficiently distribute the calculation over the GPU cores. For problems with up to a few hundred thousand non-zeros in the ILP constraint matrix, the GPU implementation is almost independent of the size of the problem because it can distribute an entire time step simulation on the cores. However, once the problem size crosses the threshold of a few hundred thousand non-zeros in the ILP constraint matrix, the simulation time of each time step of the circuit starts growing with the size of the problem. For this work we used Virtual Machines with 8-NVIDIA V100 GPUs on Google Cloud to run the GPU tests. Even if the current implementation would efficiently distribute on the GPU cores for problem sizes up to several hundreds of bits, this threshold can be increased using either larger GPUs like Nvidia A100 or H100, or other distributed architectures. 

We fully tuned the problem sizes up to 150 bits (Tuning region of figure \ref{fig:scaling_sieve_a}) and partially fine-tuned the ones at 200 bits. We stopped our tests at 300 bit problems because the limited resources from the SBIR did not allow us more cloud time to further tune or resources to finish upgrades to our CAD tools described in section \ref{subsec:Design Parameter tuning}, which would enhance the tuning efficiency. In Figure \ref{fig:scaling_sieve_b}, the scaling from the converging region is extrapolated up to 2048 bits. The timing for the ASIC realization of the MEMCPU Circuit is also reported. The ASIC timing has been estimated as in the previous section since the MEMCPU Platform, being a circuit emulator, returns the full dynamics of the circuit, including the simulated run time. It is worth noting that the forecast for the ASIC shows the possibility of solving a 2048-bit factorization problem, in sub-second time once the tuning is extended. Further note that the MEMCPU Platform running on GPUs shows the potential to factor 2048-bits in a reasonable time without the ASIC if the proper tuning is extended.

\section{\label{sec:circuit_design}Circuit design procedure}

The design of the SOGs for factorization involves several aspects, from the design of the gates and circuit topology to the parameter tuning which establishes the correct behavior of the circuit. During this SBIR, we also added extra features to the MEMCPU Platform that allowed us to enhance the design of our circuit. In this section, we will go through the most significant and impactful functionality. It is worth reiterating that the MEMCPU Platform is essentially a CAD tool to design the circuit that can also be used directly as an ILP solver. Therefore, the design with our CAD tool is also the initial step of the ASIC development.  

\subsection{Circuit architecture and layout} \label{subsec:circuit_layout}

\begin{figure*}
\includegraphics[width=0.8\textwidth]{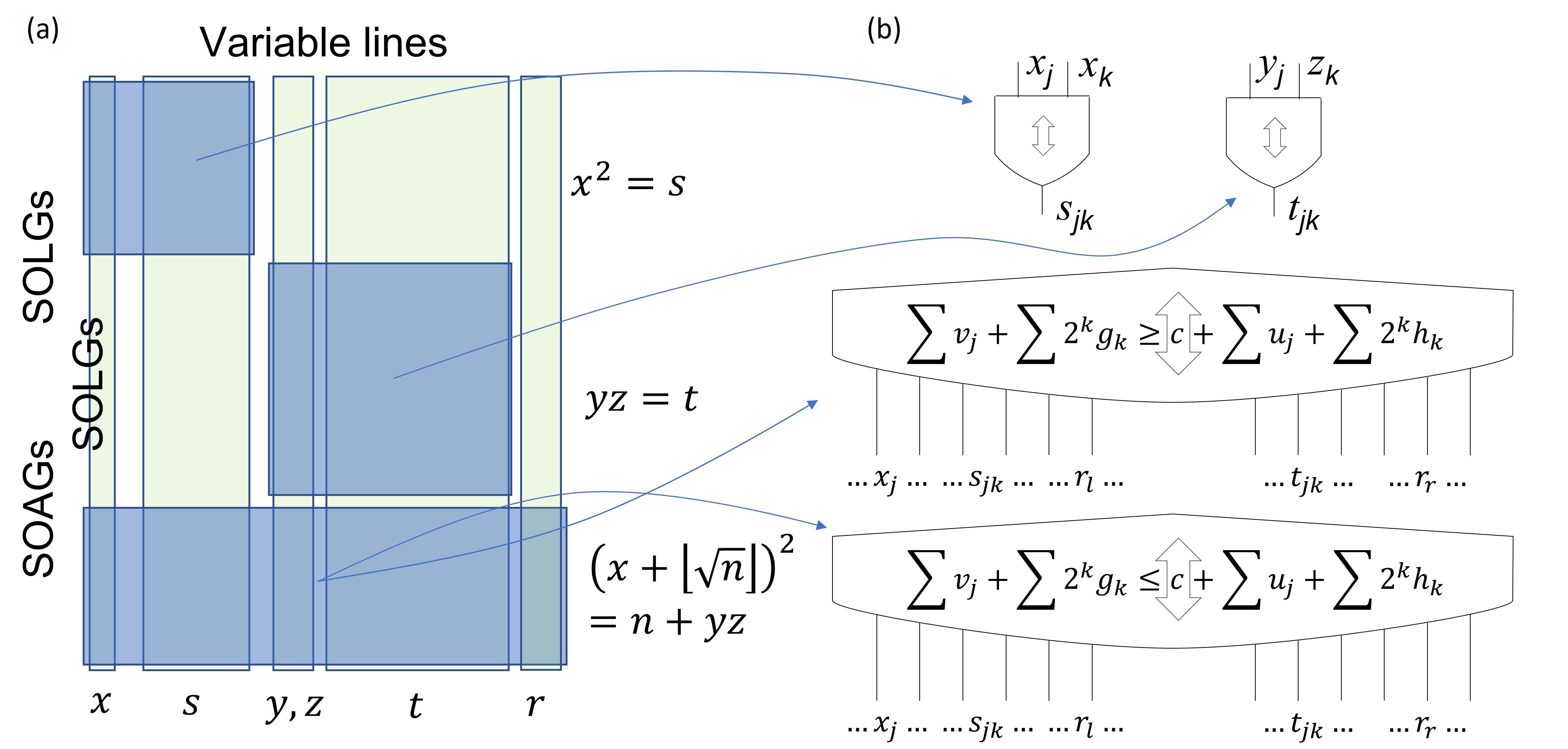} 
\caption{(a) Sketch of the MEMCPU Circuit layout implementing the ILP problem \eqref{eq:reference_equation}, with vertical bit lines and gate-containing regions. (b) Types of gates in each region. The first two regions implement the bitwise product of two integers using Self-Organizing AND gates. The bottom part of the circuit implements equations with additional terms to accurately represent the ILP problem. Equalities are implemented using a pair of SOAGs.}
\label{fig:memcpu_circuit}
\end{figure*}

The MEMCPU Platform functionality we discuss here implements the ILP problem of the quadratic form 
\begin{equation}
    (x+\lfloor \sqrt{n} \rfloor)^2=yz\label{eq:reference_equation}
\end{equation}
with $x$, $y$ and $z$ unknown integers. This equation is general enough to cover all aspects of the SOG design for all other ILP models discussed in this work. Figure \ref{fig:memcpu_circuit}a shows a sketch of the MEMCPU Circuit layout. The bit lines are vertical interconnects that traverse the circuit. For example, the light green band above $x$ indicates the interconnects, i.e., the bit lines, for the binary coefficients $x_j$ of $x$. The light blue regions indicate the areas containing gates. Figure \ref{fig:memcpu_circuit}b shows the types of gates in each region. The first two regions implement the bitwise product of two integers, as in ILP equation \ref{eq:two_inequal}. The most natural way to implement this is using Self-Organizing AND gates as depicted in Fig. \ref{fig:memcpu_circuit}b. The bottom part of the circuit implements equations of the form shown in eq. \ref{eq:prime_product}, with additional terms to accurately represent the ILP problem \ref{eq:quad_sieve}. Equalities are implemented using a pair of SOAGs, splitting an equality into two inequalities with opposite directions. From an electronic circuit perspective, these gates are compact and possible to implement using CMOS technology since their function is a linear combination of binary variables with coefficients of either 1 or a power of 2. Further details about the design are not shared as they are partly trade secrets and partly patent pending \cite{traversa_system_2022}.  

\subsection{Point Dissipativeness}\label{subsec:point_diss}

Before proceeding further with the description of the circuit design, it is important to have a better understanding of the working principles of the MEMCPU Circuit. Even if the idea of the SOGs can be perceived as relatively simple at first, the details present the challenges. This is how we make sure that the MEMCPU Circuit, i.e., a network of interconnected SOGs, converges to a state where all gates are satisfied (i.e., the solution of the problem), and does it efficiently (i.e., as quickly as possible). There are many ways to achieve these goals. Our approach is two-fold. First, we ensure convergence by looking at the point dissipativeness, a property of dynamical systems \cite{traversa_polynomial-time_2017, ventra_self-organizing_2018, hale_asymptotic_2010}. Second, we try to address the convergence efficiency by leveraging the criticality hypothesis \cite{beggs_criticality_2008} which will be discussed in the next section. 

\begin{figure}[b]
\includegraphics[width=0.47\textwidth]{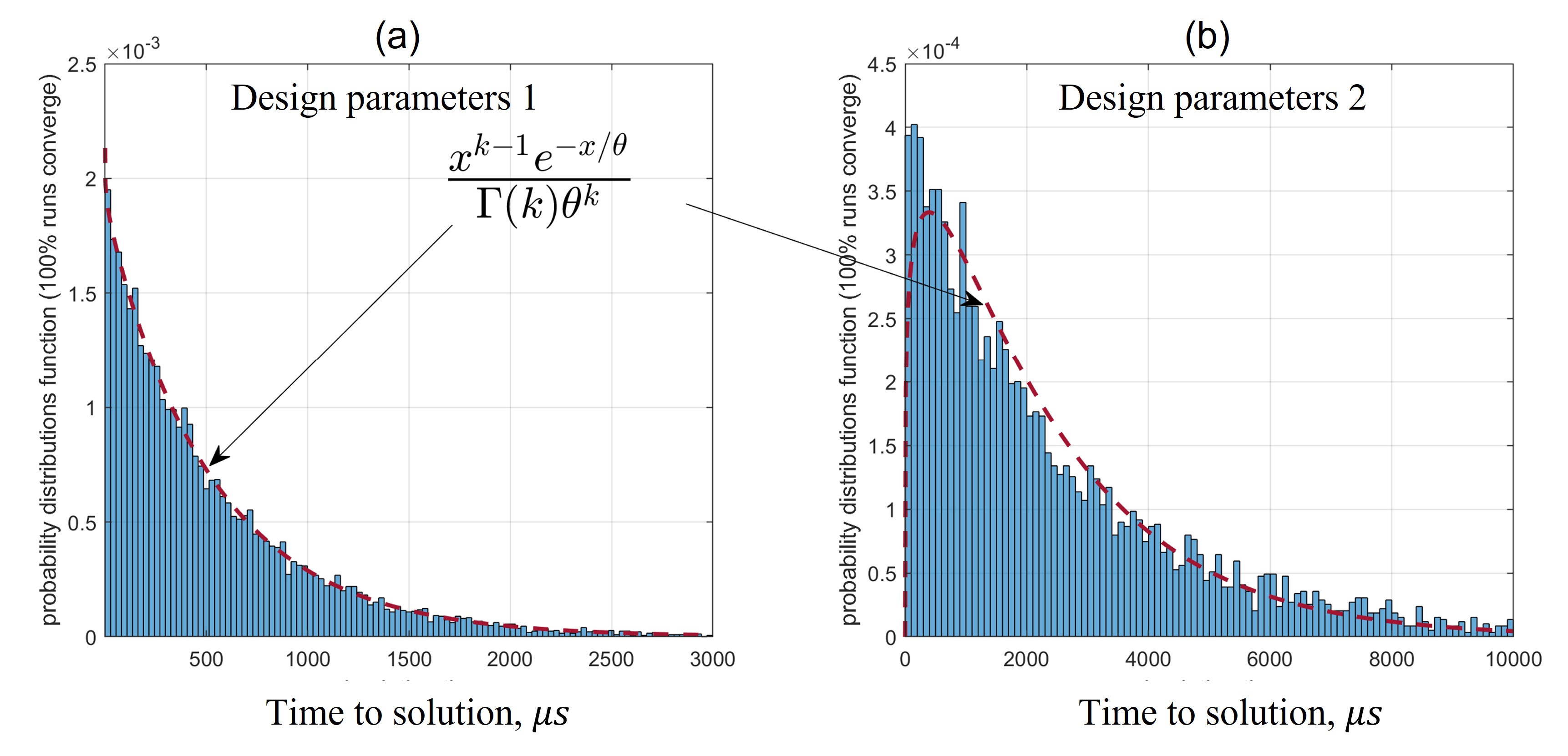} 
\caption{Distribution of time to solution for two different SOG designs in a 30-bit factorization problem, direct model. Both distributions fit well with a gamma distribution. Design 2 (b) is more than 4 times slower to converge than Design 1 (a). The peak of Distribution 1 is almost at time 0, while Distribution 2 has a peak shifted at around 500µs. This suggests that point dissipativeness is not enough in practice and that there are designs where most initial conditions lead to an almost immediate collapse to the equilibrium point of the circuit.}
\label{fig:two_design_distrib}
\end{figure}

Point dissipativeness is a property of some dynamical systems \cite{hale_asymptotic_2010} that, when translated into electronic circuit terminology, implies that the systems maintain locally bounded energy \cite{traversa_polynomial-time_2017} and have certain stability properties \cite{traversa_polynomial-time_2017,hale_asymptotic_2010}. The most important property is that the equilibria form a global attractor set \cite{traversa_polynomial-time_2017}, i.e., any system trajectory, for any initial condition, will end up into one of these equilibria. More details about this are out of the scope of this work but a great reference on the topic is \cite{hale_asymptotic_2010}. However, it is important to notice that, if this property is satisfied, then the network of SOGs will certainly reach convergence \cite{traversa_polynomial-time_2017} (that is, an equilibrium point, if it exists, in which all gates are satisfied \cite{traversa_polynomial-time_2017}) and no spurious cycles \cite{di_ventra_absence_2017} or chaos \cite{di_ventra_absence_2017-1} can emerge. While the point dissipativeness is mathematically well defined \cite{hale_asymptotic_2010}, it is hard to prove that a dynamical system satisfies it. In the case of SOGs, an isolated SOG has a design that is easy to prove to be point dissipative \cite{traversa_polynomial-time_2017, ventra_self-organizing_2018}. However, once connected in a network, it becomes challenging to prove the same for the whole system as point dissipativeness is a global property of dynamical systems \cite{hale_asymptotic_2010}. Nonetheless, it can be proved that, if a circuit made of interconnected point dissipative SOGs converges to its equilibrium point for any possible initial condition, then the entire network is point dissipative \cite{hale_asymptotic_2010}. This last statement may seem abstract, but it can be used to verify numerically that a MEMCPU Circuit design is point dissipative. In Figure \ref{fig:two_design_distrib} for example, two different design parameters for the SOGs are tested. The circuit tested in this case is for the direct factorization of a 30-bit factorization problem. In the charts, the distribution of the time to solution is shown (i.e., to the equilibrium point for the entire circuit) for random initial conditions of the circuit. We tested more than 100,000 initial conditions to produce these histograms. The simulations were stopped only if the circuit converged, and for both designs, 100\% of the simulations converged. This can be interpreted as a numerical test to prove point dissipativeness. On the other hand, Figure \ref{fig:parallel_tempering}c shows two designs for the same problem, with a more advanced circuit design that has much faster convergence. While Design 1 converges for 100\% of the initial conditions, Design 2 only converges for 42\% of the initial conditions, showing that the latter design is not point dissipative. A closer look at the distributions in Figures \ref{fig:two_design_distrib} and \ref{fig:parallel_tempering}c shows that the point dissipative designs fit very well with a gamma distribution (dashed curves). We have found that this is common for all point dissipative designs and for any problems we have tested so far. However, if point dissipativeness is not satisfied, as demonstrated by Design 2 in Figure \ref{fig:parallel_tempering}c, then the distribution largely deviates from the gamma distribution. We have found this scenario to be quite general, but we do not yet have a robust theory that explains it, although we are working on it.  

This point dissipativeness result is not at all trivial. In fact, it shows that a large network of asynchronous, autonomous, active, coupled electronic circuits can always find the trajectory to an equilibrium point, regardless of the initial conditions. This can ultimately be interpreted as the SOGs working collectively to find a common equilibrium. This is exactly what underlies the singular capabilities of the technology.
Also, this apparently general property (the time-to-solution distribution following a gamma distribution if the design is point dissipative) has many practical implications. For example, it can be leveraged to efficiently parallelize multiple initial conditions and for an efficient restart process that accelerates convergence to the solution.

\subsection{Criticality}

Point dissipativeness is not the only crucial property that we need to guarantee in our design. The speed of convergence is also crucial. A classical approach to study the speed of convergence for point dissipative systems would start from the analysis of the convergence rate of the stable manifolds employing the classical stability theory methods \cite{perko_differential_2001} applied to point dissipative dynamical systems \cite{hale_asymptotic_2010}. For complex and large dynamical systems like ours, this approach wouldn’t provide a useful outcome, both in practice and in theory, beyond some very general conclusions. 

A less standard approach is to assume the so-called critical hypothesis \cite{beggs_criticality_2008, christensen_complexity_2005}. In short, it assumes that the most efficient form of computation for a distributed system occurs at the edge of chaos, where a $2^{nd}$ order-like order-disorder phase transition takes place \cite{beggs_criticality_2008, christensen_complexity_2005}. In this scenario, at the phase transition, the system enters a critical state where long-range correlations favor scale-free properties of the system \cite{beggs_criticality_2008, christensen_complexity_2005}. This mechanism is associated with optimal information flow through the system and is thought to occur in systems like the neural cortex, making brain information processing highly efficient \cite{beggs_criticality_2008}. By applying similar hypotheses to our circuit, we aim to unravel the mechanism for optimizing computational efficiency, which ultimately results in faster convergence to equilibria.

In the past, we have analyzed critical behavior by studying long-range correlations related to instantonic processes \cite{di_ventra_topological_2017, bearden_instantons_2018}, and some of the same researchers have also explored avalanche phenomena and critical branching processes \cite{sheldon_taming_2019, bearden_critical_2019}. However, in this section, we would like to focus more on the communication aspect of the critical behaviors of memcomputing circuits. We believe it shows more promise for helping with circuit design and also provides a more practical understanding of the working principles governed by collective dynamics. The analysis reported here is not exhaustive and there is still work in progress, but it paves the way for more detailed studies and opens up new avenues for novel design techniques for SOGs.

Let’s go back to Figure \ref{fig:two_design_distrib} and take a closer look at the distributions. Both distributions fit very well with a gamma distribution (dashed curve in Figure \ref{fig:two_design_distrib}) as discussed in the previous section. Here we discuss some other implications. We first notice that the average time to solution for the two different designs in Figure \ref{fig:two_design_distrib}, is very different. On average, Design 2 is more than 4 times slower to converge than Design 1. Additionally, the peak of Distribution 1 is almost at time 0, while Distribution 2 has a peak shifted at around 500µs. This suggests that: 
\begin{enumerate}[label={(\alph*)}]
    \item Even if the system is point dissipative, we can have very different speeds of convergence depending on the design. 
	\item There are designs where most of the initial conditions lead to an almost immediate collapse into the equilibrium point of the circuit.
 \end{enumerate}

Implication a) tells us that point dissipativeness is not enough in practice. We need to find, among all point dissipative designs, those that have faster convergence. Implication b), on the other hand, highlights a much more subtle and intriguing aspect of the system. To better understand this, let’s keep in mind that this particular MEMCPU Circuit design has around 1,000 variables and a few thousand interconnected SOGs. If we then consider all electronic elements that are in the gates, this circuit has tens of thousands of state variables. Therefore, we are dealing with an autonomous dynamical system moving in a space with a dimensionality of tens of thousands. The fact that such a large dynamical system can collapse into its equilibrium point in just a few time steps for most random initial conditions is not common or trivial, especially for systems that are networks of autonomous dynamical systems. 

\begin{figure}[b]
\includegraphics[width=0.47\textwidth]{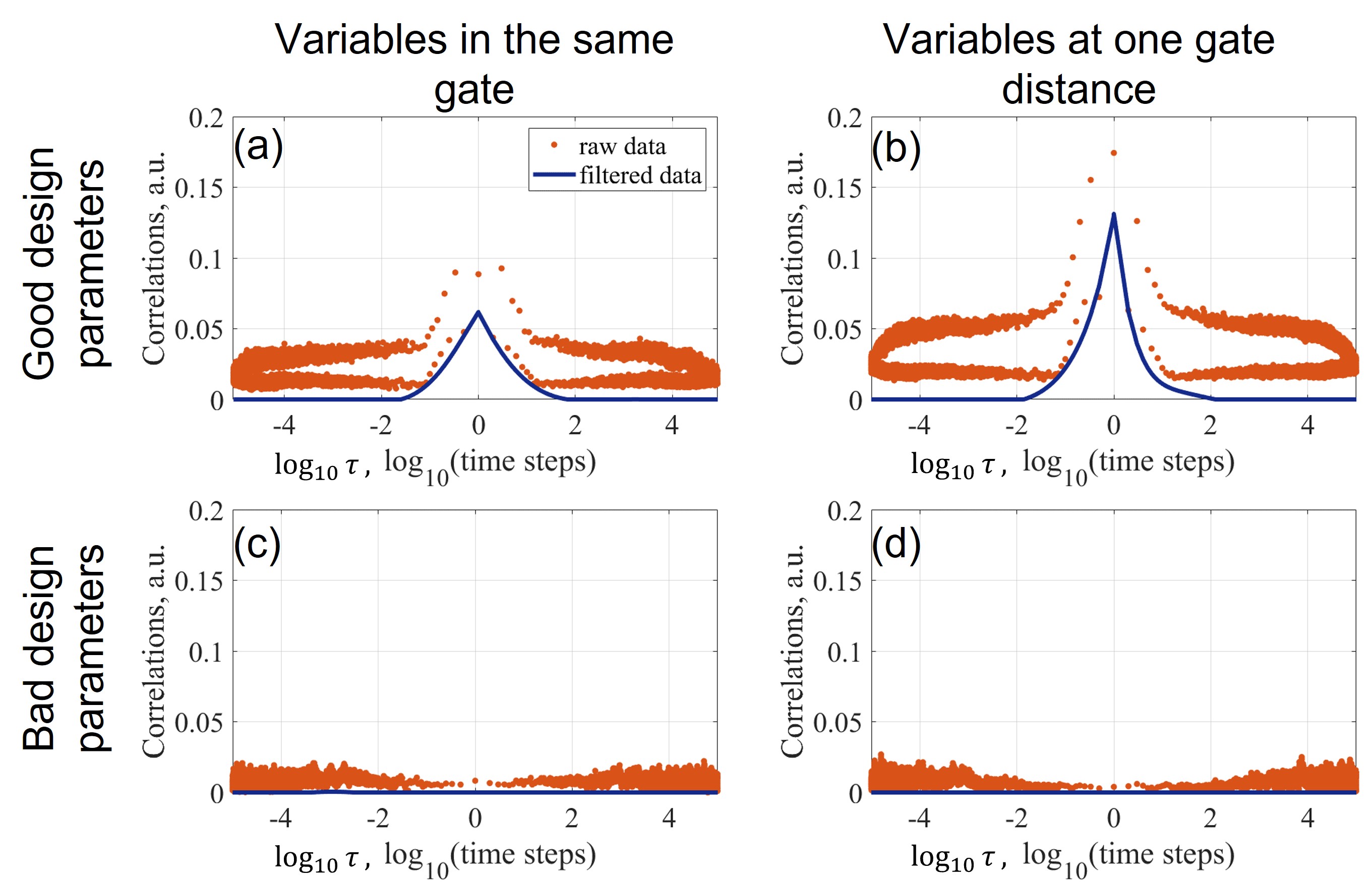} 
\caption{Visualization of internal correlations for two different SOG designs. The first design ((a) and (b)) was point dissipative with a very good convergence speed, while the second design ((c) and (d)) did not converge to equilibrium within the allotted simulation time. The good design exhibits high correlations that persist for up to order-one-hundred time steps, while the bad design does not show any correlation above the noise level. Figures (a) and (c) show correlations for pairs of bit lines that are both connected to at least one SOG, while Figures (b) and (d) show correlations of bit lines that require at least one SOG to be traversed to go from one line to the other. For good designs, correlations extend beyond those internal to the gates and non-locally to the entire circuit, creating effective mutual communication among gates.}
\label{fig:correlation}
\end{figure}

The reasons for this behavior can be found looking at the correlations that are established in the system and what they imply. In Figure \ref{fig:correlation} for example, we show the correlations between the threshold crossing of the voltage for two different pairs of circuit nodes and two different designs. The correlations are defined as $C_{jk} (\tau)=R_{jk} (\tau)/\sqrt{(R_{jj} (0) R_{kk} (0))}$ where $ R_{jk}=\int f(v_j (t))f(v_k (t+\tau)) dt $, $v_j$ is the voltage at the bit line $j$ and $f$ is the threshold crossing function defined as 
\begin{align*}
f(x) = \begin{cases} 1 & \text{if } \max(x(t-\Delta t), x(t)) > th \quad \wedge \\ &\min(x(t-\Delta t), x(t)) < th \\ 0 & \text{otherwise} \end{cases}
\end{align*}
for a threshold $th\in R$. These correlations indicate whether two SOG terminal voltages are crossing the threshold (and therefore switching their binary state) due to communication established by the SOGs. The correlations in Figure \ref{fig:correlation} are for circuits emulated for $10^5$ time steps. We have reported the raw correlations calculated using the convolution theorem and filtered correlations using standard procedures such as polynomial fitting and baselining to the noise level.
Visualizing internal correlations is a powerful analysis tool for SOGs. Figure \ref{fig:correlation} highlights two different designs. The first design (the good one) was tested for point dissipativeness and has a very good convergence speed (Figure \ref{fig:correlation}a and \ref{fig:correlation}b). It is similar to the design in Figure \ref{fig:two_design_distrib}a. The second design (the bad one) could not converge to equilibrium within the allotted simulation time, regardless of the initial conditions (Figure \ref{fig:correlation}c and \ref{fig:correlation}d). It is important to note that both designs are for a network of interconnected point dissipative SOGs. However, as discussed in the previous section, point dissipativeness does not necessarily extend to the network as a global property of the dynamical system. Looking at correlations, we see that the good design exhibits high correlations that persist for up to a few hundred time steps. In contrast, the bad design does not show any correlation above the noise level, demonstrating no useful communication among nodes. Another important aspect is that the correlations are quite symmetric (i.e., for positive and negative $\tau$), indicating that the good design has no preferential flow of information within the circuit and communication is mutual. Although Figure \ref{fig:correlation} only shows correlations for two pairs of circuit nodes, we checked most of the node pairs and all showed similar behavior.

Figure \ref{fig:correlation}a and \ref{fig:correlation}c show correlations for pairs of bit lines that are both connected to at least one SOG. On the other hand, Figure \ref{fig:correlation}b and \ref{fig:correlation}d show correlations of bit lines that require at least one SOG to be traversed to go from one line to the other. For the MEMCPU Circuit layout depicted in section \ref{subsec:circuit_layout}, it is unlikely for more distant lines in terms of SOGs to be traversed since large SOAGs, like the ones in Figure \ref{fig:memcpu_circuit}, are connected to many bit lines and the average distance in terms of SOGs to be traversed is 1. Therefore, this second node pair represents non-local correlations in terms of SOGs. Figure \ref{fig:correlation}b shows that for good designs, correlations extend beyond those internal to the gates and non-locally to the entire circuit, creating effective mutual communication among gates.

In conclusion for this subsection, we can summarize that the convergence process is driven and accelerated by mutual communication among bit lines happening in and among the SOGs. This communication has non-local properties, as mutual correlations persist even if the bit lines are not directly electronically coupled through an SOG. The high persistent correlations, their mutual and non-local nature, are all fingerprints of critical behavior happening in the circuit. Our interpretation of critical behavior indicates that the circuit is able to have gates communicating at long distances, and the communication is mutual and balanced. Since we found these global features always for designs whose tests show point dissipativeness, we conclude that this criticality-driven communication is likely the mechanism that, once enhanced, accelerates the convergence to the equilibria. There is additional work in progress to quantify this mechanism and uncover critical parameters that can be used to enhance the design of our circuits.

\subsection{Design Parameter tuning}\label{subsec:Design Parameter tuning}

An interesting theoretical question at this point is: What is the ideal design for SOGs and what would it imply? Answers to this question can be found in both classical stability theory of autonomous systems \cite{perko_differential_2001} and complexity and criticality theory \cite{christensen_complexity_2005}. However, this goes beyond the scope of this work. A more practical question is: How we can design a MEMCPU Circuit that approaches the ideal design. To this end, over the last few years, we have developed methods that are still being perfected and implemented within our in-house Computer-Aided-Design (CAD) platform. The design process can be summarized as follows:
\begin{enumerate}[label={(\alph*)}]
	\item Each SOG is designed with a set of free parameters that can vary within a certain range. The design guarantees, within the parameter range, point dissipativeness of each gate if isolated. These parameters are typical design parameters for electronic components, ranging from resistances and capacitances to transistor model parameters.
	\item We input the ILP model into our CAD platform using standard formats such as .mps or .lp files. During the SBIR performance period, we upgraded the CAD platform to allow the input of extra information about the problem structure, as well. Specifically, subgroups of SOGs and SOG terminals can be demarcated as belonging to different groups that we call “families”. This allows for the independent design of multiple families of components within the circuit.
	\item We use an internal general-purpose optimization method to explore the parameter space. It is a heavily modified parallel tempering Monte Carlo method \cite{earl_parallel_2005} with the goal of quickly returning design parameters that could potentially make the entire circuit point dissipative and approach criticality. This method is implemented in our CAD platform and is used to explore the parameter space for the different SOG families.
	\item We further analyze the candidate designs from parallel tempering to filter out those that are point dissipative, have the best convergence, and exhibit the best generalizability to other problem instances. 
	\item This process is repeated at increasing problem sizes to scale up to larger ones. Using a continuation algorithm to accelerate the design process, we use designs from the previous sizes as starting points for the larger ones.
 \end{enumerate}
 
\begin{figure*}
\includegraphics[width=0.95\textwidth]{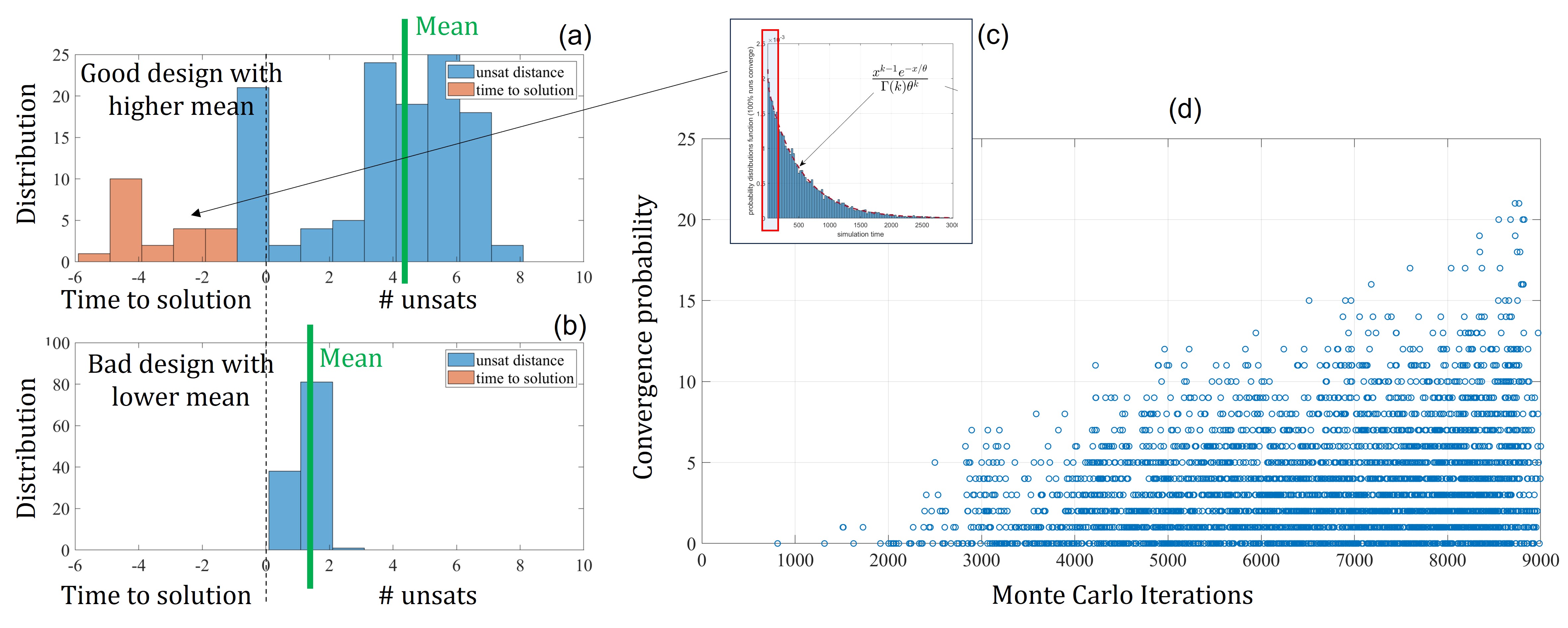} 
\caption{Comparison of time to solution, unsats and convergence probability for different SOG designs. (a) Distribution of the number of unsats for the good design, which remains at high values before quickly reaching the solution. (b) Distribution for a bad design that can quickly arrive at a few unsats but never succeeds at converging. (c) Distribution of time to solution for a good design, with the part of the distribution that would be observed within a short simulation time highlighted by the red box. (d) A typical run of parallel tempering using an objective function (\ref{eq:objective_fn}) for a 38-bit problem direct model, with the probability of convergence for short simulations increasing as the number of iterations increases.}
\label{fig:time_distribution}
\end{figure*}
As we mentioned, this process is still being perfected, even though it already provides reasonable results. To give an example of the challenges that we face, let us discuss one of the most important ones that were the subject of intense development during the SBIR. We would like the parallel tempering to return parameter designs that make the circuit point dissipative and have high convergence speed, which is highly critical. However, even though we are working on it, as of today we do not have a parallel tempering objective function that directly measures these two properties. So, we use an objective function that indirectly optimizes towards these two goals, which poses serious challenges. For example, the most basic idea would be to use some distance function defined for the SOGs that remain unsatisfied during the simulation. The simplest approach would be to just count the unsatisfied SOGs at each time step during the simulation and use the minimum attained. We can call this the number of “unsats”. Even though this may seem like a good idea at first, let us describe the severe problems it generates.   

The biggest issue arises from the fact that, since parallel tempering explores a large and complex space, it needs many iterations to reach a good region. An iteration is a short simulation to test the design parameters of each chain. The number of iterations varies depending on the number of SOGs families and the number of Monte Carlo chains we use, but it usually ranges between a few thousand and tens of thousands of iterations to achieve reasonable results. To give an idea of the size of the parameter space, let us consider the design in Figure \ref{fig:memcpu_circuit}. We can define several families of variables and SOGs. A basic classification would be 4 variable families $(x,[s,t],[y,z]$, and $r$) and 3 SOG families (the SOANDs for $x^2=s$, the SOANDs for $yz=t$, and the SOAGs), which, it turns out, would translate into 121 free design parameters to be optimized. The parameters are not generally independent (obviously, since being point dissipative and critical are global properties), so the parallel tempering must explore the entire space of 121 dimensions for continuous parameters. For a size like this, a good number of iterations for our parallel tempering would be a few tens of thousands with at least 180 chains divided into 6 independent parallel tempering routines of 30 chains each. This implies that long circuit simulations would be prohibitive, and therefore we test parameter designs with just short simulations. 

Since we have a distribution of time to solution depending on the initial conditions, “short simulations” means that even good designs may not have enough time to arrive at the solution during an iteration of parallel tempering. For example, let’s consider again the parameter set in Figure \ref{fig:two_design_distrib}a, which is also reported in Figure \ref{fig:time_distribution}c for convenience. Typically, for factorization, we simulate about 100 µs of circuit dynamics. In Figure \ref{fig:time_distribution}c, the part of the distribution that we would observe within such short time is highlighted by the red box. Many initial conditions would end with a positive number of unsats. A distribution of the number of unsats can be found in Figure \ref{fig:time_distribution}a. It is interesting to note that this parameter (already known to be point dissipative and sufficiently critical) has several unsatisfied clauses during most of the simulation. It remains at high values before finally quickly reaching the solution, collapsing into equilibrium. Turning to bad designs, there are many poor circuit designs that can quickly arrive at a few unsats but never succeed at converging. This is seen for example for the circuit design that produced the distribution in Figure \ref{fig:time_distribution}b. These are designs in which some elements in the circuit may not have enough energy to trigger the correct dynamics, resulting in metastable states that screen the equilibrium points. Unfortunately, this means that using unsats (the number of unsatisfied SOGs) can lead parallel tempering into the wrong region, resulting in designs that are unlikely to be point dissipative.         

We found a mitigation to this problem by using the following objective function for parallel tempering,
\begin{align}\label{eq:objective_fn}
    \text{obj} = \mu -\sqrt{\sigma^2}+\sqrt[3]{m_3}, 
\end{align}
where $\mu=\braket{\text{score}}$, $\sigma^2= \braket{(\text{score}-\mu)^2 }$, $m_3= \braket{(\text{score}-\mu)^3 }$, and score is the value of any distance function that quantifies how unsatisfied the SOGs are. The objective function of Eq. \ref{eq:objective_fn} corrects the mean of the score (which, as explained earlier, can be very misleading) with the standard deviation and the third-order moment to statistically quantify the probability of reaching a solution within a short test. However, implementing this objective function in parallel tempering can be challenging because it requires an estimate of the aforementioned expectation values. Typically, we test a design only once and then accept or reject it based on the Metropolis-Hastings algorithm \cite{earl_parallel_2005} before moving on. To overcome this issue, we estimate \ref{eq:objective_fn} by using previously tested design parameters close to the ones under testing on the current iteration. Using statistical inference, this method provides reasonable results, especially if the probability of convergence for short runs and good designs is above 5-10\%. Figure \ref{fig:time_distribution}d shows a typical run of our parallel tempering using \ref{eq:objective_fn} as the objective function for a 38-bit problem, for the direct factorization method, and a total of 82 free parameters. For designs that converged during parallel tempering, we also post-processed the probability of convergence for short simulations by rerunning them with 100 different initial conditions and timing out at 100 µs of circuit dynamics. Figure \ref{fig:time_distribution}d shows that as the number of iterations increases, this probability also increases, indicating that parallel tempering is converging to design parameter regions where the circuit designs are increasingly efficient in converging to their equilibria and thus solving the factorization problem. 
\begin{figure*}
\includegraphics[width=0.95\textwidth]{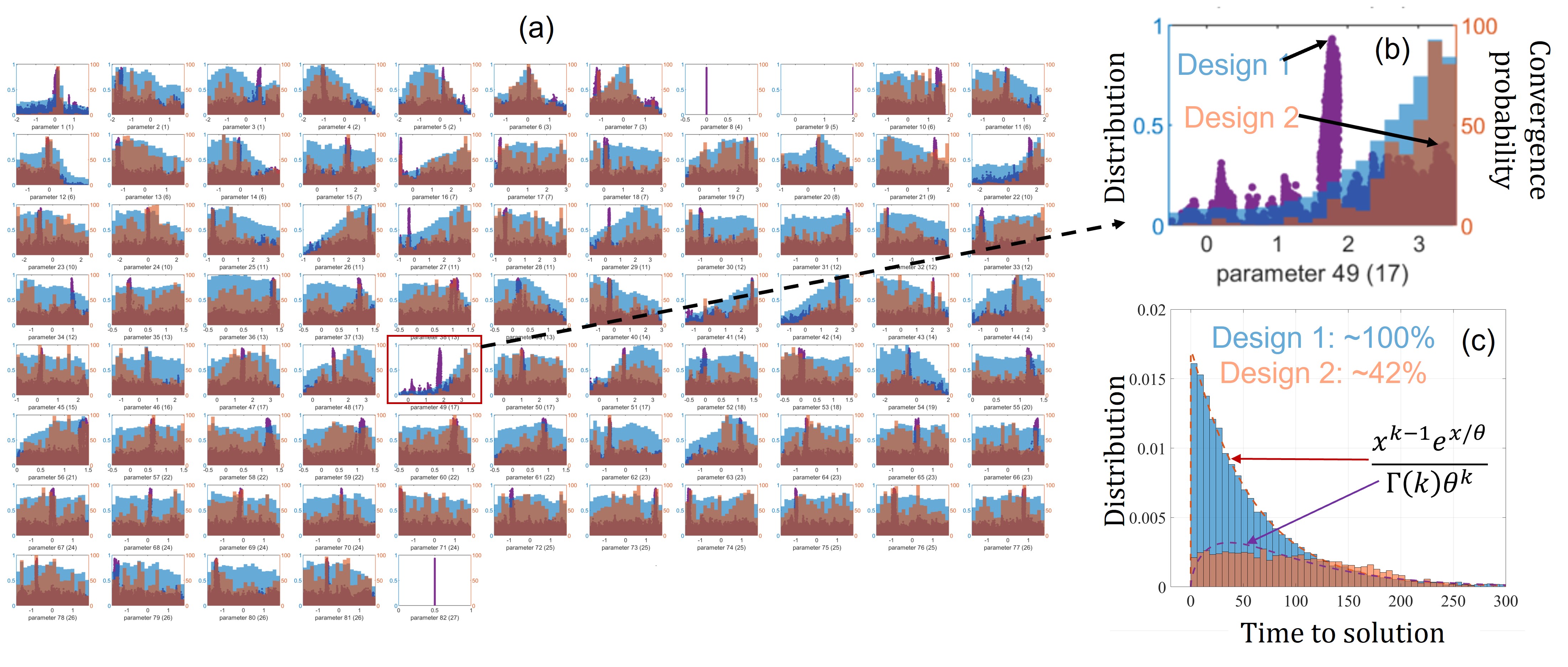}
\caption{Distributions of the parallel tempering iterations in the parameter space.(a) Sampling distribution of parallel tempering using \ref{eq:objective_fn} as the objective function. (b) Distribution for parameter 49. The design 1 parameters come from the peak at 2 and the design 2 parameters are taken from the edge as shown. (c) Convergence rates for sample designs from the peak of the distribution and from the region where parallel tempering was most attracted show very different behaviors.}
\label{fig:parallel_tempering}
\end{figure*}

Unfortunately, even this method is not perfect, as parallel tempering can still be attracted to regions of the design parameter space where the circuit is not point dissipative. An example is shown in Figure \ref{fig:parallel_tempering}. Figure \ref{fig:parallel_tempering}a shows the sampling distribution of our parallel tempering using \ref{eq:objective_fn} as the objective function for a 30-bit problem, the direct factorization method, and a total of 82 free parameters, with 180 chains divided into 6 independent parallel tempering runs, each running for 12,000 iterations. Each subplot reports the histogram of the trial samples (light blue), the samples that converged to equilibrium during the short run (light orange), and the post-processed probability of convergence for short runs (100 µs of circuit dynamics) for the design samples that converged during parallel tempering. To understand the issue, let us concentrate on just one free parameter, parameter 49, as shown in Figure \ref{fig:parallel_tempering}b. Here, we can see that the projection of the distribution shows a peak at around 2. If we test some sample designs from that peak, we find that they are point dissipative and have very good convergence rates and high and symmetric correlations, indicating critical behavior (Figure \ref{fig:parallel_tempering}c). However, the parallel tempering did not spend much time in that region, as the sample distribution suggests. On the other hand, if we test a parameter from the peak of the sample distribution, i.e., from the region where parallel tempering was most attracted, we see that the design associated with it is not point dissipative, showing only 42\% of maximum convergence as reported in Figure \ref{fig:parallel_tempering}c and also discussed in Section \ref{subsec:point_diss}. This demonstrates that even though we use \ref{eq:objective_fn} to get some good designs that provide the scaling reported in Section \ref{subsec:mem_approach}, the parallel tempering objective is still not perfect, and more work needs to be done to arrive quicker at ideal designs. We are currently exploring the use of an upgraded objective based on a critical parameter related to correlations that can be used to correct  \ref{eq:objective_fn} and drive parallel tempering into regions where designs exhibit high correlations and are also point dissipative.

Although the effort to tune is currently somewhat onerous, we usually tune one problem instance per size, and then the design obtained is used for any other problem instances of the same size. For this reason, the total amount of compute for tuning was not quantified here and the computing burden was also mitigated by leveraging old gate designs to find new ones using continuation techniques. Such techniques that leverage design parameters from smaller sizes on larger sizes were seen to be very effective for tuning efficiently as size increases. The amount of tuning required is expected to continue with the same trend and at this point we do not see any roadblock that cannot be overcome with proper R\&D and compute resources.

\begin{figure}[b]
\includegraphics[width=0.35\textwidth]{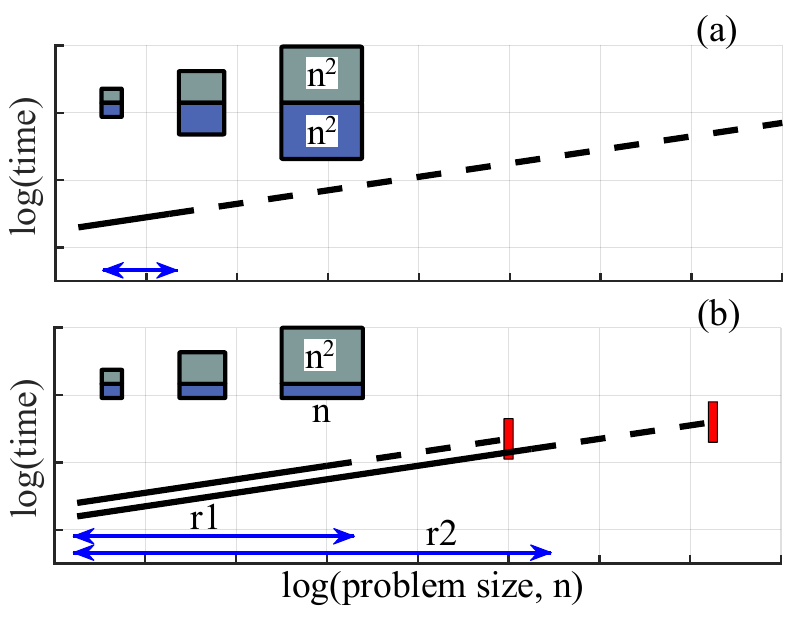}
\caption{Schematic illustrating that problems with size invariant structure can be tuned at a single problem size  (solid line, tuning region), expecting to use the tuned parameters at all sizes  (dashed line, neighboring region) (a). The inset depicts a circuit layout with two types of gates that grow at the same rate. In contrast, the structure of factorization changes with size (b), \emph{e.g.} changing the ratio of product gates to sum gates. Tuning on example problems of a small range (r1) of sizes (solid line, tuning region) is seen to produce convergence not only for other problems in the range, but also up to a size somewhat larger than r1 (dashed line, neighboring region). Tuning further (r2) is seen to continue the scaling (red bar).
}
\label{fig:schematic_retuning}
\end{figure}

Finally, to summarize the effect of design parameter tuning, in Figure \ref{fig:schematic_retuning} we illustrated how the effect of the tuning extends beyond the tuning region. If the problem we are dealing with roughly maintains a constant structure in terms of node voltage connectivity, proportion of types of gates, terminals per gate, etc, then by tuning for only small problem sizes, the obtained design parameters will maintain the same circuit and convergence properties (point dissipativeness and criticality) at higher problem sizes. Therefore what we defined as the neighbouring region extends to any size. On the contrary, if the structure changes with the problem size, then the neighbouring region will extend only partially and to maintain the same circuit and convergence properties, we have to tune at higher and higher sizes as depicted in Figure \ref{fig:schematic_retuning}. This is also remarked in sections \ref{subsubsec:direct_model} and \ref{subsubsec:sieve_model}.

\subsection{\label{sec:transtion_asic}Forecasted results from ASIC implementation}

In the last few sections, we introduced and discussed the working principles of the MEMCPU Circuit and how our in-house CAD tool is used to optimize circuit design. Currently, our CAD tool implements generic compact models for electronic devices. 
The MEMCPU Platform results correspond to the time to perform the simulation of these generic components. How much time passes \emph{within} the simulation provides the forecasted ASIC results. This is only an estimate of the future ASIC results, though, because the generic components (\emph{e.g.} resistors, capacitors, transistors) currently simulated must be substituted for foundry models, making any necessary accommodations, before layout generation and manufacturing.

\section{Conclusion}

In this SBIR, we explored both direct and congruence approaches for prime factorization. We found that the MEMCPU Platform achieved best acceleration using the congruence method. Our variant of the sieve method is based on the same principles as the quadratic sieve or GNFS methods. However, it accelerates the solution by using the MEMCPU Platform to return unique congruences, i.e., special relations among integers, that are ultimately used to factorize large biprimes (numbers that have exactly two prime factors)  \cite{pomerance_smooth_2008}. 

To use the MEMCPU Platform as if it was a sieve machine to find these congruences, we have developed an integer linear programming (ILP) formulation whose solutions are congruences. Therefore, it is sufficient to solve the ILP multiple times to factorize the biprime. Sieve-based methods work similarly, but conventional sieving has super-polynomial complexity \cite{lenstra_development_1993, pomerance_smooth_2008}, as shown in Figure \ref{fig:scaling_sieve}. Figure \ref{fig:scaling_sieve} also demonstrates that a best-in-class ILP solver, like Gurobi, does not reduce the complexity (as expected) since it is not efficient in solving the ILP problem to find congruences. However, a properly designed memcomputing circuit provides congruences very efficiently at a rate that is well described by a $2^{nd}$ order polynomial fit as a function of the problem size, as reported in Figure \ref{fig:scaling_sieve}. Our approach also allows us to control the smoothness of the congruences, such that we can keep the number of congruences bounded with the size of the problem, as discussed in Section \ref{subsubsec:sieve_model}. This is not the optimal choice for standard sieve methods where they need an optimal trade off between smoothness and operations to find smooth congruences leading to their well known sub-exponential scaling.

To understand the challenges, a quick review of the MEMCPU Platform was provided. The MEMCPU Platform emulates a circuit made of Self-Organizing Gates (SOGs) \cite{traversa_polynomial-time_2017, ventra_self-organizing_2018}. Each SOG is an electronic component whose voltages at the terminals encode variables of the problem and the SOG drives these voltages to satisfy a relation among the variables \cite{traversa_polynomial-time_2017, ventra_self-organizing_2018}. We call it “self-organizing” because it drives voltages autonomously via feedback loops and, once the SOG satisfies the feedback relation, it shuts down. When SOGs are connected via their shared terminals, the feedback from each SOG is propagated through the entire circuit. Hence, the circuit works in unison to satisfy all SOGs at once to minimize the feedback \cite{traversa_polynomial-time_2017, ventra_self-organizing_2018}. However, it is important that energy balance and signal propagation properties be satisfied in order for the circuit to work properly \cite{di_ventra_topological_2017, sheldon_taming_2019}. We have provided specific details about these design aspects through our analysis of point dissipativeness and criticality, which characterize the working principle of the MEMCPU Circuit. Additionally, we have discussed how we use and tune design parameters to ensure these properties.

During the SBIR, we were able to find these parameters for designs that can handle up to 300 bit factorization problems. It is worth noting that these parameters are size dependent and not problem instance dependent. Continuing further requires finding parameters that scale this forward using continuation methods as discussed in this report.

\begin{acknowledgments}
We gratefully acknowledge the financial support provided by the USAF Intelligence group through the SBIR Phase II, under Contract number FA864922P1131. Our sincere thanks to Dr. Massimiliano Di Ventra for his valuable insights and for reviewing this manuscript. We also extend our gratitude to Dr. Sergio Decherchi for his stimulating discussions.
\end{acknowledgments}

\appendix

\section{Industrial applications}\label{app:industrial_application}

\subsection{Oil and Gas Maritime Delivery Scheduling }

\begin{figure}[b]
\includegraphics[width=0.4\textwidth]{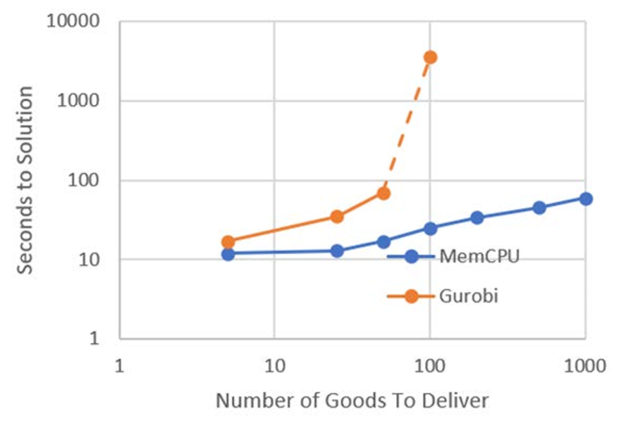}
\caption{Performance comparison of the MEMCPU Platform and Gurobi in solving an ILP problem for scheduling goods delivery in the Oil \& Gas industry, showing linear scaling for the MEMCPU Platform and exponential scaling for Gurobi as the number of goods increased on a log-log scale.}
\label{fig:oil_gas_compare}
\end{figure}

Here is an example application of a complex problem in the Oil \& Gas industry \cite{traversa_oil_2020}. This work involved developing an optimized 30-day schedule for delivering 3500+ goods from a port in Louisiana to, from, and between offshore oil platforms in the Gulf of Mexico. The corporation employed several ships with various capacities for different types of goods. In addition, the ships also deliver fuel to the platforms and must ensure they do not run out of fuel. The goal was to provide a highly optimized shipping schedule while minimizing the number of ships and prioritizing slow over fast ships, since fast ships have limited capacity and are much more expensive to run. This type of scheduling problem in its full size is usually not amenable to classical optimization techniques, so companies manually create schedules based on simple operational rules, such as a first-come, first-served approach. 

This problem can be cast as ILP, transforming all scheduling constraints into linear inequalities invoking binary, integer, and continuous variables. The resulting problem is quite heterogeneous and complex but includes every aspect of the problem without approximations. Addressing the 40 different types of constraints results in hundreds of thousands of linear inequalities and hundreds of thousands of variables of various types. The resulting ILP is extremely hard, but its optimal solution will eliminate the inefficiencies of the first-come, first-served approach. The MEMCPU Platform and Gurobi (Fig. \ref{fig:oil_gas_compare}) were evaluated side-by-side. To perform a scaling analysis, we created multiple problems with increasing numbers of goods to be distributed. Gurobi, as expected, showed rapid exponential scaling, demonstrating the combinatorial hardness of the problem. At 100 goods, Gurobi’s computational time exceeded 8 hours to find any feasible solution to the problem. The MEMCPU Platform showed linear scaling as the number of goods increased (Fig. \ref{fig:oil_gas_compare}). The MEMCPU Platform could be run over the full problem (3624 goods over 30 days).

\begin{table*}%
\caption{\label{tab:oil_gas_results}%
Results of using the MEMCPU Platform to optimize the delivery schedule, showing a decrease in the number of required ships, an increase in goods delivered by slow ships, a reduction in goods required to be delivered by expensive fast ships, and a decrease in the total number of transits, resulting in significant cost savings and reduction in greenhouse gases.
}
\begin{ruledtabular}
\begin{tabular}{l|lll|lll}
&
\# Slow Ships &
\# Transits &
\# Goods &
\# Fast Ships &
\# Transits &
\# Goods\\
\hline
 Actual & 12 & 176 & 2899 & 2 & 32 & 725\\
MEMCPU & 7 & 92 &3472 & 2 & 10 & 152\\
Improvement & 41\% less & 48\% less & 20\% less & same & 68\% less & 79\% less
\end{tabular}
\end{ruledtabular}
\end{table*}

The resulting schedule provided the wide array of benefits listed in (Table \ref{tab:oil_gas_results})
These optimizations should then deliver the following benefits:
\begin{itemize}
    \item \$0.5M monthly (\$6M annual) reduction in ship lease payments.
    \item \$1M monthly (\$12M annual) reduction in fuel costs.
    \item 18kt reduction in greenhouse gases.
    \item Reduction in ship maintenance costs, crew costs, overtime, etc.
    \item These savings can then be replicated worldwide.
\end{itemize}

This work demonstrated that the MEMCPU Platform can efficiently solve combinatorial optimization problems considered intractable for today’s best-in-class solutions while delivering significant efficiency improvements.

\subsection{Aircraft, Passenger, Cargo, and Crew Scheduling}

MemComputing has addressed several commercial aircraft scheduling problems. The Defense Innovation Unit (DIU) challenged us with an airlift scheduling problem specifically for the military \cite{traversa_aircraft_2019}. They provided simulated airlift data representing worldwide military aircraft, passenger, cargo, and crew scheduling. The problem statement is quite complex, and details can be found in \cite{traversa_aircraft_2019}. However, it can be summarized as follows. 
\begin{itemize}
    \item \textbf{Airlift requests}: There is a collection of airlift requests. Each of these requests a group of passengers and/or cargo. Each request can be fulfilled using multiple aircrafts distributing passengers and cargo. Each airlift has a priority code, departure location, earliest pickup date, latest pickup date, latest drop-off date, destination, number of passengers, number of pallets of cargo and constraints on the type of aircraft. 
    \item \textbf{Aircraft fleet}: The fleet consists of two types of aircraft. Each aircraft identifies the current airfield, the capacity for passengers and pallets of cargo, the cruising speed, and the maximum hours of continuous flight. There are also regulations that identify when, where and for how long an aircraft is unavailable for maintenance or other issues. 
    \item \textbf{Crew}:  There are regulations that must be met for the safety of the crew. These include the maximum number of hours the crew can be active, minimum down time after completing their active day and minimum time on the ground between legs. 
    \item The goal is to create a schedule that successfully books all priority-1 requests with as many priority-2+ requests as possible while minimizing the overall flight time. Passengers and cargo may require multiple legs to reach their final destinations. These can be satisfied with the same aircraft or by switching aircraft as is most optimized. 
\end{itemize}

\begin{figure}
     \subfloat[\label{fig:aircraft_a}]{%
  \includegraphics[width=0.8\columnwidth]{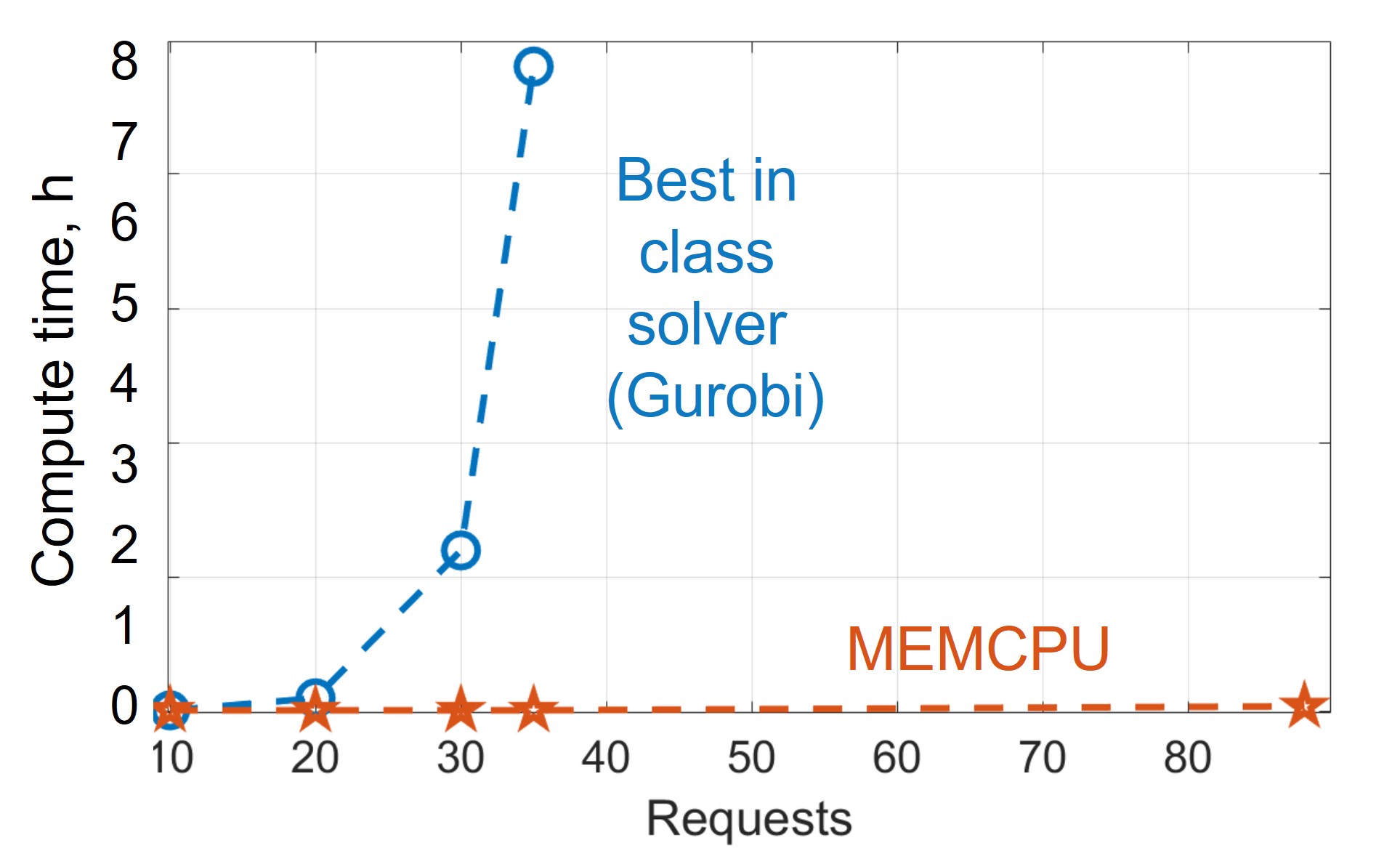}%
}
\\
\subfloat[\label{fig:aircraft_b}]{%
  \includegraphics[width=0.8\columnwidth]{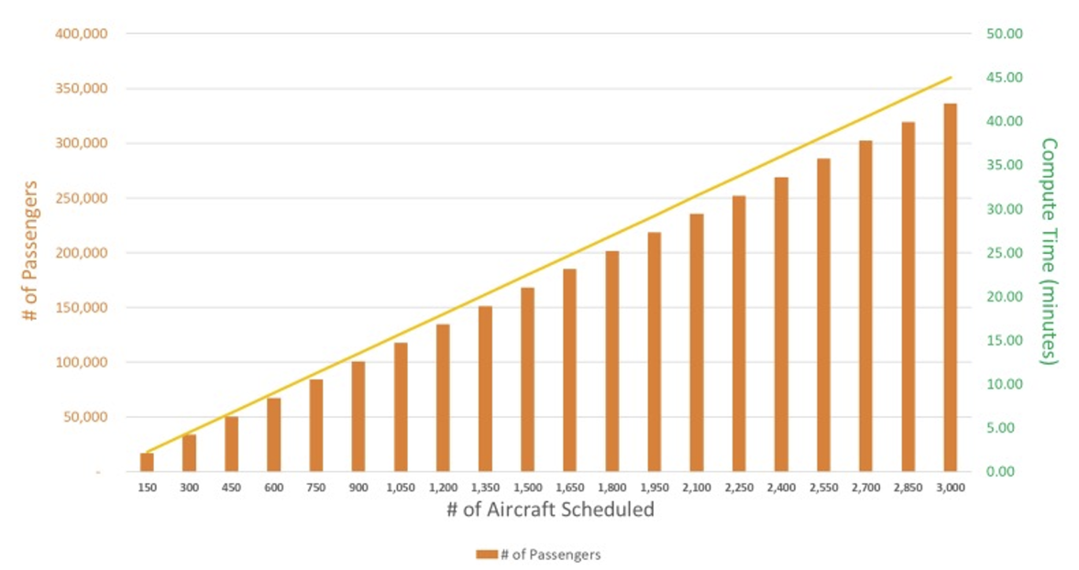}%
}
\caption{(a) Performance comparison between the MEMCPU Platform and Gurobi in solving an ILP problem for military airlift scheduling, showing linear scaling for the MEMCPU Platform and exponential scaling for Gurobi as the number of airlift requests increased. (b) Results of using the MEMCPU Platform to optimize airlift schedules, demonstrating the ability to automate, optimize, and help large commercial airlines recover quickly from major flight interruptions, resulting in significant cost savings in fuel and efficiency.}
\label{fig:aircraft}
\end{figure}

We cast this problem in ILP format without the need for any approximations. In Fig. \ref{fig:aircraft_a}, a performance comparison between the MEMCPU Platform and Gurobi, a best-in-class ILP solver, is shown when solving for varying numbers of airlift requests. Gurobi could solve small problems quickly, but its performance degraded exponentially as the number of airlift requests increased. On the other hand, the MEMCPU Platform solved all instances within minutes, exhibiting linear scaling. For example, it took approximately 1.5 minutes to calculate an optimized schedule for the full problem presented by the DIU that covered 2-weeks of airlift requests.
This approach can scale much higher as depicted in Fig. \ref{fig:aircraft_b}. There we tested scenarios of a scale like the meltdown of Southwest Airlines in December of 2022 \cite{traversa_aircraft_2019, dunn_southwest_2022}. In summary, the work demonstrated that our approach can automate, optimize and help large commercial airlines recover quickly from major flight interruptions that can be brought on by large weather fronts. Savings in fuel and efficiency from fast, highly optimized scheduling can exceed hundreds of millions annually for these companies as well the DOD \cite{josephs_southwest_2023,traversa_aircraft_2019}.

\subsection{Drone Swarm Optimization}

Lockheed Martin presented us a head-to-head benchmark study comparing the MEMCPU Platform vs. the best-in-class solution used to find the optimal path for a swarm of drones moving in a complex environment like a maze.

\textbf{Problem Statement}: A set of agents must move in an environment with obstacles, like a maze. Each agent has its own mission where it must travel from its starting point to its ending point. The goal is to find the optimal path for all agents that avoids deadlocks, collisions, or other interference \cite{traversa_drone_nodate}. 
This is a well-known problem in literature and goes under the name of Multi Agent Path Finding  (MAPF) problem \cite{traversa_drone_nodate}. The  MAPF problem is known to be exponentially difficult with the number of agents, and the best-in-class solver used today is based on the A* optimization algorithm \cite{traversa_drone_nodate}. 

\begin{figure}
     \subfloat[\label{fig:drone_a}]{%
  \includegraphics[width=0.8\columnwidth]{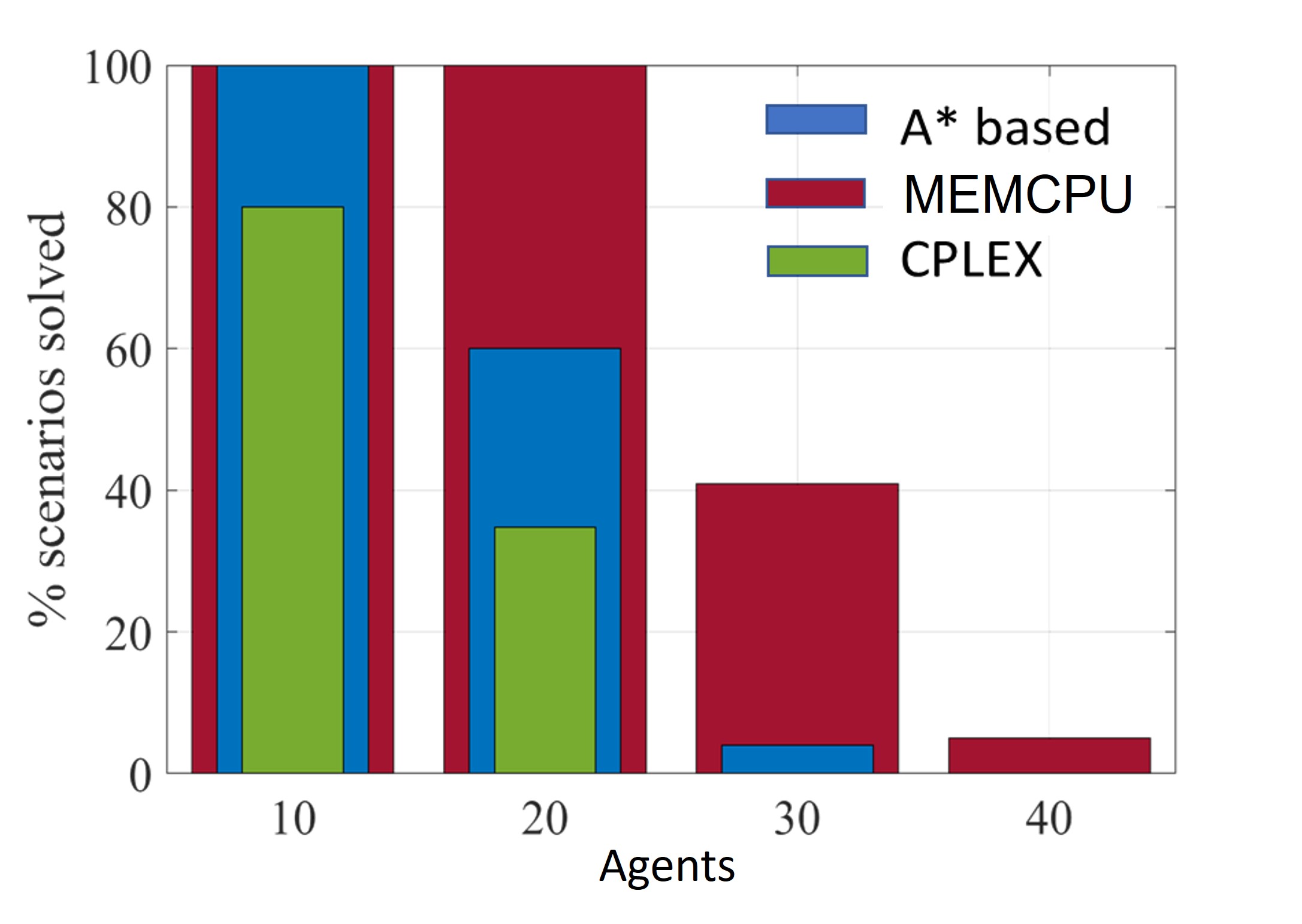}%
}
\\
\subfloat[\label{fig:drone_bc}]{%
  \includegraphics[width=0.8\columnwidth]{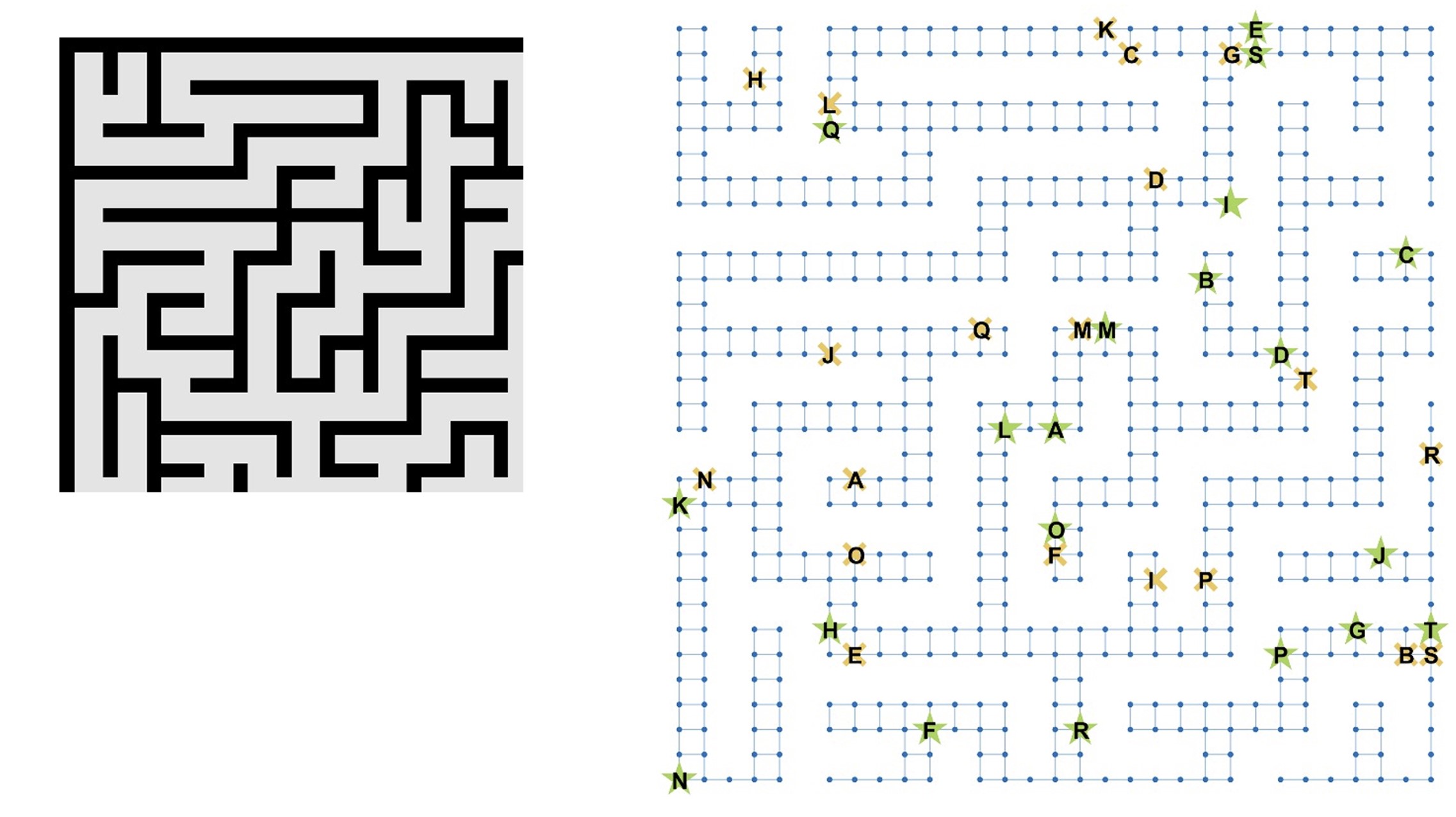}%
}
\caption{(a) Results of a head-to-head benchmark study comparing the MEMCPU Platform, CPLEX, and an in-house method based on the A* algorithm in solving a Multi Agent Path Finding (MAPF) problem for a swarm of drones moving in a complex environment, showing that the MEMCPU Platform provided solutions to many more scenarios as the number of agents increased. (b) The maze used in the challenge (left) and the graph representation of the maze, with symbols representing the starting and ending points for a set of agents (right).}
\label{fig:drone}
\end{figure}

The head-to-head challenge that Lockheed Martin presented used the maze from Fig. \ref{fig:drone_bc} (left). The maze was represented in the form of a graph (Fig. \ref{fig:drone_bc}, right) so that trajectories of agents are described by indicating when they transit the nodes, and starting and ending points for all agents are constrained. Using graph theory, we developed an ILP formulation representing the MAPF problem. The ILP was suitable for both the MEMCPU Platform and CPLEX, which was run by Lockheed Martin. Lockheed Martin also used an in-house method based on the A* algorithm, considered to be the most efficient method known for MAPF problems. All solvers were given 10 minutes to find solutions. The MEMCPU Platform outperformed the other two solvers, providing solutions to many more scenarios especially as the number of agents increased. A summary of the final results is reported in Fig. \ref{fig:drone_a}.
\bibliography{references}

\end{document}